    \documentclass[final,numbers,5p,times,twocolumn]{elsarticle}

   \usepackage{graphicx}
   \usepackage{epsfig}

\usepackage{amssymb}

   \biboptions{longnamesfirst,comma}

\journal{             }

\begin{document}

\begin{frontmatter}



\title{Pulsar-driven Jets in Supernovae, Gamma-Ray Bursts, 
and the Universe}


\author{John Middleditch}

\ead{jon@lanl.gov}
\address{MS B265, Los Alamos National Lab., Los Alamos, NM 87544}

\begin{abstract}
The bipolarity of Supernova 1987A can be understood 
through its very early light curve observed from the CTIO 
0.4-m telescope and IUE FES, and following 
speckle observations of the `Mystery Spot' by two groups. These 
indicate a highly directional beam/jet of light/particles,
with initial collimation factors in excess of 10$^4$ 
and velocities in excess of 0.95 c, involving 
up to 10$^{-5}$ M$_{\bigodot}$ interacting with circumstellar 
material.  These can be produced by a model proposed in 1972, by
Bolotovskii and Ginzburg, which employs pulsar emission from 
polarization currents induced/(modulated faster than c) beyond the 
pulsar light cylinder by the periodic electromagnetic field, 
(supraluminally induced polarization currents -- SLIP). 
SLIP accounts for the disruption of progenitors in supernova 
explosions and their anomalous dimming at cosmological 
distances, jets from Sco X-1 and SS 433, the lack/presence 
of intermittent pulsations from the high/low luminosity low mass 
X-ray binaries, long/short gamma-ray bursts and predicts that 
their afterglows are the {\it pulsed} optical/near infrared emission 
associated with these pulsars. SLIP may also account for the 
TeV e$^+$/e$^-$ results from PAMELA and ATIC, the WMAP `Haze'/Fermi 
`Bubbles', and the r-process. SLIP jets from SNe of the first 
stars may allow galaxies to form without dark matter, and explain the 
peculiar, non-gravitational motions observed from pairs of distant 
galaxies by GALEX. 

\end{abstract}

\begin{keyword}
acceleration of particles \sep plasmas \sep gamma rays: bursts 
\sep pulsars: general \sep stars: neutron \sep supernovae: general 
\sep supernovae: individual (SN 1987A) \sep ISM: jets and outflows 
\sep dark matter \sep cosmology: observations \sep distance scale 
\sep early universe


\end{keyword}

\end{frontmatter}


\section{Introduction}
\label{intro}

\begin{figure}
   \centering
   \includegraphics{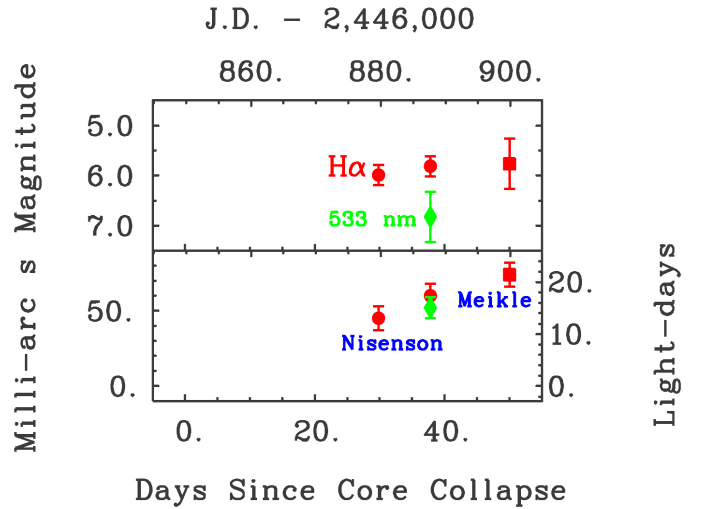}
   \caption{Measurements of displacement (lower) and observed
   magnitude (upper) of the `Mystery Spot' from SN 1987A,
   at H${\alpha}$ and 533 nm, vs time, from [1-3].
               }
              \label{magdist}%
    \end{figure}

Supernova 1987A has provided astronomers with a wealth of data,
some of which has not even now, a quarter century after the event, 
been satisfactorily accounted for by any model.
One of the most remarkable features
in the early study of SN 1987A was the `Mystery Spot',
with a thermal energy of 10$^{49}$ erg, observed 50 days after
the core-collapse event [1-3], and separated from the SN photosphere
`proper' by 0.060$\pm$0.008 arc s at day 38 (Fig.~\ref{magdist}), with about 3\%
of this energy eventually radiated in the optical band.
The possibility that the enormous energy implied for the Mystery Spot might
somehow link it to gamma-ray bursts (GRBs) attracted little serious
consideration at the time, or even since, beyond a very astute few
[4-6].  The Mystery Spot was also observed at separations
of 0.045$\pm$0.008 arc s on day 30, and 0.074$\pm$0.008 arc s on day
50, but always at an angle of 194$^{\circ}$, consistent with the southern
(and approaching) extension of the bipolarity \citep{Wa02}.  The Mystery
Spot offsets from SN 1987A imply a minimum projected separation of $\sim$10
light-days ($\ell$t-d).

There is a also wealth of photometric and spectroscopic data from even
earlier stages of SN 1987A, in particular photometry data from the
Cerro Tololo Inter-American Observatory (CTIO) 0.4-m telescope \citep{HS90},
and the Fine Error Sensor (FES) of the International Ultraviolet Explorer
(IUE -- \citep{Wa87}), and spectroscopic data from
\citep{Men87} and \citep{Da87}, among others.
Short timescale structure ($\le$1 d) in this data, following finite
delays ($\sim$10 d) after SN 1987A core-collapse, implies at least
one beam of light and jet particles which were initially highly collimated
($>$10$^4$), interacting with circumstellar material.

GRBs, particularly long, soft GRBs ($\ell$GRBs), appear to be the most
luminous events in the Universe, occurring at the SN rate of one per
second (assuming a collimation factor near 10$^5$) yet we still know
very little about them (see, e.g., \citep{Mz06}, and references
therein).  A few $\ell$GRBs have been found to be associated with SNe
[13-15].  Others however, mostly those with slightly harder spectra and 
lasting only $\sim$1 second, (sGRBs), usually produce no lingering evidence 
at all, and only rarely the `afterglows' observed for a large proportion of 
$\ell$GRBs, sometimes extending down to radio wavelengths.  A large 
number of models have been put forth to explain GRBs, including neutron
star-neutron star mergers
for sGRBs, and exotic objects such as `collapsars' \citep{MW99} for
$\ell$GRBs.  Part of the physical motivation for these, and the nearly
universal use of their usually severe degrees of collimation, is the 
enormous energy of up to 10$^{54}$ erg implied for an isotropic source.
Indeed the data from SN 1987A presented herein, when interpreted through the 
model discussed in detail within this work, support a
beam/jet collimation factor $>$10$^4$ in producing its
early light curve by interaction with more-or-less stationary
circumstellar material,
but in addition predicts that GRB afterglows, and by extension of the
principles of supraluminal excitations, even GRBs themselves, may dim 
only as distance$^{-1}$ (see $\S$\ref{link}), and thus easily removes 
the requirement for such a high energy.

Specifically this work argues that polarization currents,
induced beyond the light cylinders of, and
by the rotating magnetic fields from,
newly-formed pulsars embedded within their stellar remnants
\citep{KPR74,BR01} can account for the bipolarity of SN 1987A 
[19-21].
This model of emission from supraluminally induced polarization currents
(SLIP) provides a mechanism for generating a pulsed beam, rotating on the
surface of a cone whose half angle (and angle from the pulsar
axis of rotation) is given by,
\begin{equation}
\theta_{\rm V} = \arcsin {\rm c}/v,
\end{equation}
for astronomical distances.  Here c is the speed of light,
and $v>{\rm c}$ is the speed at which the
polarization currents are updated, i.e., $v = \omega R$,
where $\omega$ is the pulsar rotation frequency, in radians s$^{-1}$,
$R>R_{\rm LC}$ is the distance of the polarization current from the 
pulsar, projected onto the rotational equatorial plane, and $R_{\rm LC}$ 
is the light cylinder radius ($\omega R_{\rm LC} = {\rm c}$).

Emission from circularly supraluminal polarization currents is 
initially focused as a single point, which moves around the equator
of the light cylinder at the angular pulsar rate, but which 
eventually travels outward on the two cones defined by Eq.~(1).  
When the emission is observed for the conal directions the pulse 
profile will consist of only a single, sharp peak, and will have 
an intensity that decays only with the first power of distance 
\citep{Ar07,SSM09} (see $\S$\ref{link}).\footnote{Because not all 
pulsars show the characteristically sharply peaked pulse profiles 
which indicate that they are observable because they obey the 
distance$^{-1}$ law, some pulsars are indeed observed from an angle, 
$\theta \ne \theta_{\rm V}$.}

This emission mechanism, and its unusual distance law, will operate 
at {\it any} radius beyond the light cylinder where sufficient plasma 
exists.  In principle this would render pulsations nearly always 
visible from neutron stars surrounded by plasma.  However, pulsations 
are almost never seen from neutron stars which are born inside of SNe 
with M$_{\odot}$'s of material yet to eject, and those in the 
non-transient (and more luminous) Low Mass X-ray Binaries (LMXBs). 
What happens is that emissions from smaller annuli must push out 
through the densest material of the stellar core or circumpulsar 
accretia, while emissions from larger annuli, although encountering 
more material on the way to the equatorial light cylinder, also bypass 
much of the core material while traveling toward the poles.
This results in 
a greater beam concentration near the rotational poles, 
and thus the solid angle for which pulsations are observable 
from SNe and LMXB's is tiny, and are almost never {\it 
expected} to be seen (a SN Ic, however, is worth observing -- see 
$\S$\ref{Ia/c}).  In the case of SNe, the beam accelerates (and
transmutes) matter from the stellar interior, which includes H, He, 
C, O, and heavier elements for massive, solitary progenitor stars. 

In the rest of this paper, $\S$\ref{early} includes a quantitative
discussion of the SN 1987A early luminosity history and motivations
for why a later, quasi-steady, less collimated, as well as a prompt,
highly collimated, injection event, is also needed.  Then $\S$\ref{geom}
estimates the kinematics and a working geometry for the 1987A beam/jet
and Mystery Spot, and we explore the implications of these on the process 
which gave rise to SN 1987A, and observations during the following years.
Section \ref{link} relates the SN 1987A beam/jet process to GRBs,
germinates the idea that their afterglows are highly pulsed,\footnote{No
observation of any GRB afterglow has yet been made in high time 
resolution, but they are expected to be highly pulsed because of
the associated inverted distance law (see$\S$ \ref{link}).  However,
the afterglows of GRBs 990510, 990712, and 020405 were measured to be 
1.7\%, 2.9-1.2\%, and 10\% linearly polarized [25-28], which could be 
a signature of pulsed emission.  The prompt emission of GRB 041219 
was also found to be highly, and variably linearly polarized by 
{\it INTEGRAL} \citep{Got09}, though we don't expect such emission
to be pulsed.}
and introduces the possibility that 1987A itself was the result of 
a merger of two electron degenerate stellar cores within a common
envelope -- double degenerate, or DD -- while
Section $\S$\ref{Ia/c} relates the process to Type Ia/c SNe.  Section
\ref{Details}, including Subsections \ref{TypeII} - \ref{HistR},
extend the discussion of the implications of the DD process
to the formation and luminosities of Type Ib and II SNe, the relation 
of plasma available to the SLIP process and transient signals 
(\ref{Plasma}), Sco X-1 and other Low Mass X-ray Binaries (LMXBs -- 
\ref{LMXBs}), historic SNe, pulse profiles, and SNR bipolarity (\ref{HistR}). 
Subsection \ref{0537} discusses how SLIP explains why PSR J0537 is 
predominantly an X-ray, but not an optical pulsar.  Subsection 
\ref{Positrons} discusses the high kinetic energy of 
the particles in the SN 1987A jet(s), and its possible 
relation to the positron excess, WMAP `Haze', and Fermi `Bubbles', 
followed by a discussion of the unique SS 433 jets in
$\S$\ref{SS433}, 
pulsar-driven jets in 
the early Universe in $\S$\ref{earlyU}, and pulsar 
`recycling' in $\S$\ref{Recycling}.  Section \ref{conc} 
concludes, and a short SLIP primer follows in the Appendix.  

   \begin{figure}[h!]
   \centering
   \includegraphics{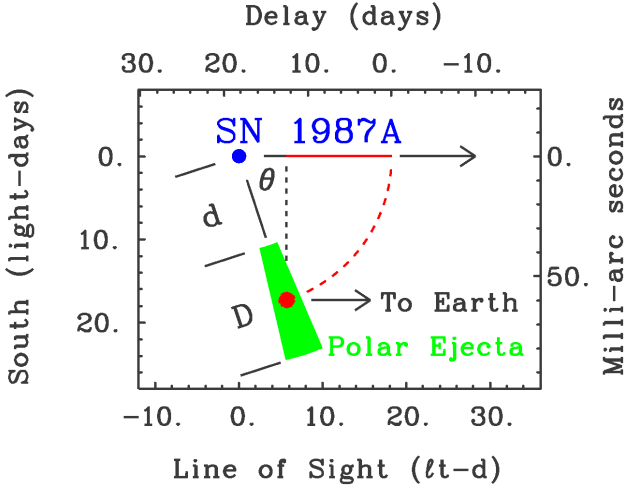}
   \caption{
   The geometry of the `Mystery Spot,' (red dot)
   polar ejecta, associated beam/jet, and direct line of sight
   from SN 1987A to the Earth.
               }
              \label{fgeom}%
    \end{figure}

\section{The Early Luminosity History of SN 1987A}
\label{early}

\begin{figure}
   \centering
    \includegraphics{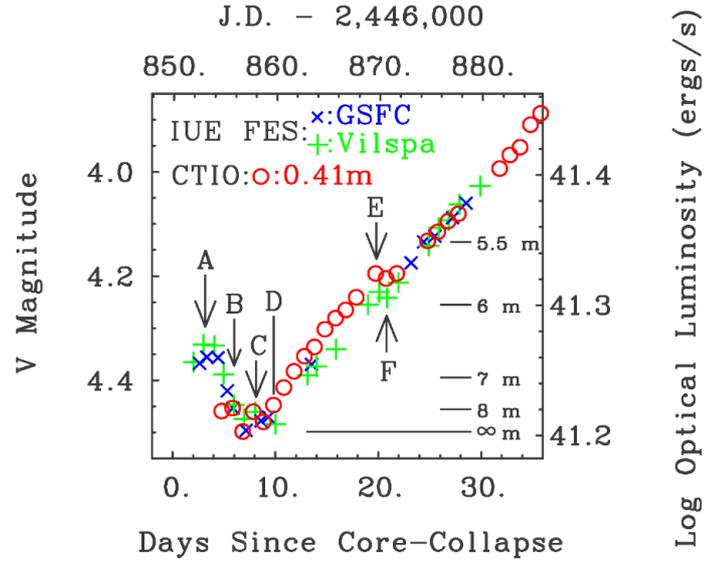}
   \caption{
The very early luminosity history of SN 1987A as observed
with the Fine Error Sensor of IUE and the 0.41-m at CTIO.  Data
points taken at Goddard Space Flight Center by Sonneborn \& Kirshner,
and at the Villafranca Station in Madrid, are marked (see
$\S$\ref{early}).  The diameter of the circles representing the
CTIO data is about three times the error.  A V magnitude scale,
for excess light above the minimum near day 7, is given, within the
frame, on the right.
               }
              \label{fearly}%
    \end{figure}

Table 1 gives an event history for SN 1987A and its
progenitor system.
An approximate geometry for SN 1987A and a Mystery Spot located
within circumstellar material (in this case, polar ejecta), is
given in Fig.~\ref{fgeom}, while the early luminosity
histories of SN 1987A from CTIO, and the Fine Error Sensor
(FES) of the IUE, which we can use to refine the parameters that
this figure postulates as reflecting reality, are both plotted in
Fig.~\ref{fearly}.\footnote{The CTIO V band
center occurs at 5,500 \AA, as opposed to 5,100
\AA ~for the FES, and in consequence, the FES magnitudes have
been diminished by 0.075 in Fig.~\ref{fearly} to account for the resulting
luminosity offset.  In addition, comparison of the CTIO data \citep{HS90} 
with the FES data at the minimum near day 7 and the decrement near day
20 reveals that the published CTIO times may be too early by
1 day, and have been adjusted in this work.  All other 
significant disagreements between the fluxes can be 
explained by temporary spectral features appearing in one band and not
in the other.} 

\begin{table}[!h]
\caption{SN 1987A Event Log}            
\label{table:1}     
\begin{tabular}{l l l }       
\hline\hline                 
Time      &&Event \\    
 t                 \\    
\hline                        
-20,000 years &&Rings formed \\ 
$\sim$(-2,000?) years &&Polar, or near-polar ejection \\
 0 (= UT 1987, Feb. 23.316) && Core-collapse of SN 1987A \\
 0$<$t$<$2 days  &&UV flash from SN 1987A  \\
 2$<$t$<$4 days  &&Emergence of luminous jet \\
 4$<$t$<$7 days  &&Cooling/spreading of jet \\
 7.8 days    &&UV flash hits polar ejecta\\
 8.26 days   &&Jet impacts polar ejecta (PE) \\
 19.8 days     &&Pulsations clear through PE \\
 20.8 days     &&Jet particles clear through PE \\
 30 days     &&`Mystery spot' at 45 mas  \\
 38 days     &&`Mystery spot' at 60 mas  \\
 50 days     &&`Mystery spot' at 74 mas  \\
\hline                                   
\end{tabular}
\end{table}

\begin{figure}[!h]
   \centering
    \includegraphics{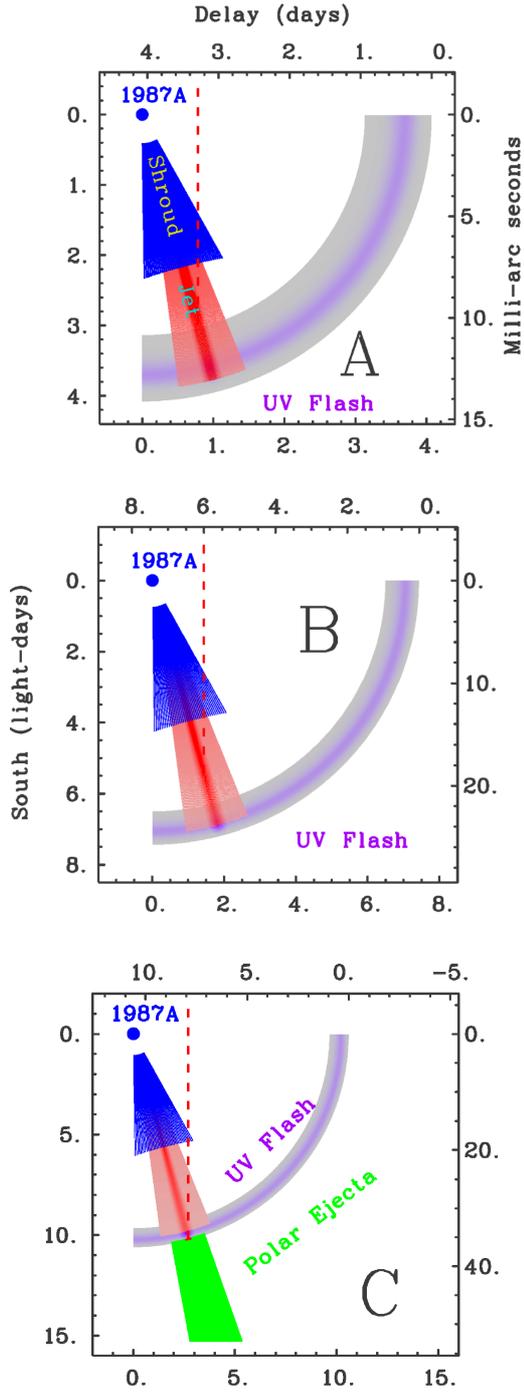}
   \caption{
The geometry of the 1987A glowing beam/jet, initially opaque shroud,
and UV flash (which may have an enhanced beam of its own in the jet
direction -- here 75$^{\circ}$, down and to the right).  
The maximum velocity of the jet/shroud is 0.95/0.55 c. 
The dashed line to the upper scale flags the center of the emerging
jet at day 3.3 (top -- A), and day 6 (middle -- B), and the UV flash
hitting the polar ejecta at day 7.8 (bottom -- C).
               }
              \label{ABC}%
    \end{figure}

Following the drop from the initial flash, the luminosity rises
again to a maximum (`A' in Fig.~\ref{fearly} and Fig.~\ref{ABC},
top) of magnitude 4.35 at day 3.0, roughly corresponding to
2.7$\times$10$^{41}$ erg s$^{-1}$
and interpretable as a luminous jet emerging from cooler,
outer layers which initially shrouded it.  This declines
to magnitude 4.48 around day 7.0 (`B', Fig.~\ref{ABC}, middle), interpretable
as free-free cooling, or the loss of the ability to cool, as the
jet becomes more diffuse.  The next observable event should be
the scattering/reprocessing of the initial UV flash in the polar
ejecta at day $\sim$8, and indeed `C' (Fig.~\ref{ABC}, bottom)
shows
$\sim$2$\times$10$^{39}$ erg s$^{-1}$ in the optical for a day at
day 7.8, and a decline {\it consistent with the flash} after that,
indicating that no significant smearing over time had occurred in
this interaction.  The 8 day delay to, and the rapid decline of, 
this first event implies a collimation factor $>$10$^4$ for this 
part of the UV flash.  This can be seen if we let $\delta$ represent
the radius of the illuminated area in the polar ejecta, $d$ as in
Fig.~\ref{fgeom}, the distance from the SN to this area, the collimation factor, 
$\Upsilon$ for the beam can be expressed as:
\begin{equation}
\Upsilon = 2\pi~({d \over \delta})^{2};
\end{equation}
or, 
\begin{equation}
\Upsilon = 10^4~({d \over 10~{\rm \ell t-d}})^2~({\delta \over 0.25~{\rm \ell t-d}})^{-2}.
\end{equation}

\begin{figure}[!h]
   \centering
    \includegraphics{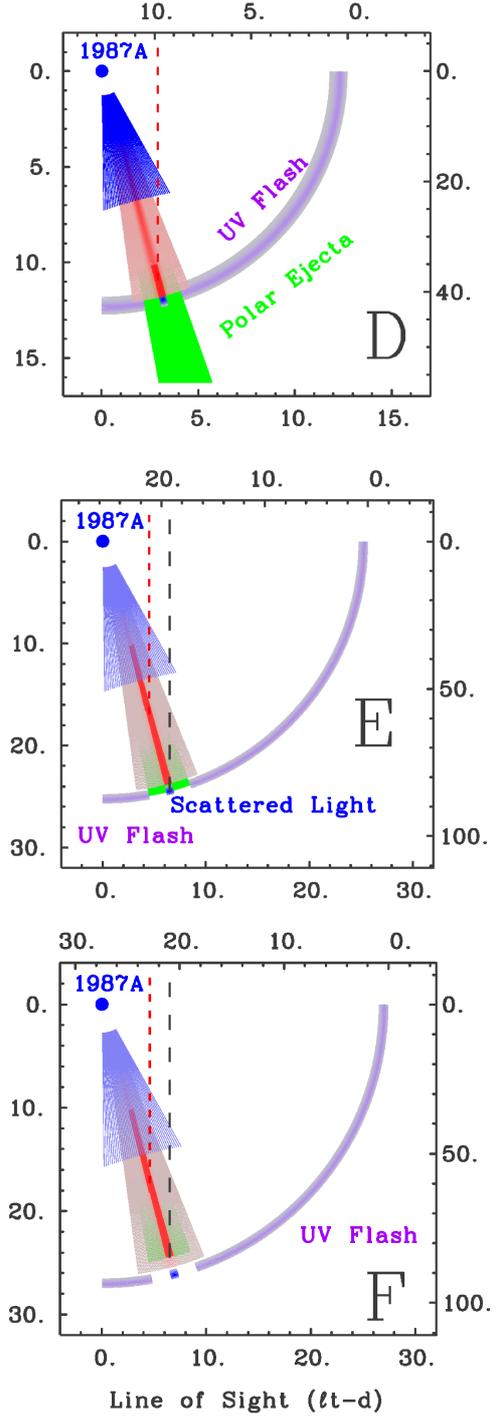}
   \caption{
(Top -- D) The intense center ($\sim$1$^{\circ}$) of the jet begins to
produce light as it penetrates into the polar ejecta, producing
the jump in luminosity at day 9.8.
(Middle -- E) Particles in the jet continue to impact the polar ejecta
(mostly hidden), extending the ramp in luminosity visible in
Fig.~\ref{fearly} near day 20 (left dashed line to the top scale).
(Right dashed line) Light from the filtered UV flash scatters in
clumpy polar ejecta near its outer boundary.  (Bottom -- F) The fastest
jet particles have cleared the end of the polar ejecta.
               }
              \label{DEF}%
    \end{figure}

A linear ramp in luminosity, visible by day 9.8 ('D' in Fig.~\ref{fearly}
and Fig.~\ref{DEF}, top),
indicates particles from
the jet penetrating into the polar ejecta, with the fastest traveling at
$>$0.8 c, {\it and}, because of the sudden rise, a particle collimation
factor $>$10$^4$.
Back-extrapolation of the three CTIO points just after day 8
intersects the day 7 minimum near day 8.26, which would indicate
that particles exist in this jet with velocities up to 0.95 c,
and even higher if the true minimum flux is lower than the points
at magnitude 4.48 (1.6$\times$10$^{41}$ erg s$^{-1}$) near day 7.

The ramp continues until after day 20, when a decrement
of $\sim$5$\times$10$^{39}$ erg s$^{-1}$ appears in both
data sets just after day 20 (`F' in Fig.~\ref{fearly}, and
Fig.~\ref{DEF}, bottom).
The CTIO point just before
the decrement can be used as a rough upper limit for the Mystery Spot
luminosity, and corresponds to an
excess above the minimum (near day 7.0) of 5$\times$10$^{40}$ erg
s$^{-1}$, or magnitude 5.8, the same as that observed in H${\alpha}$
for the Mystery Spot at days 30, 38, and 50, about 23\% of the total 
optical flux of 2.1$\times$10$^{41}$ erg s$^{-1}$ at that time.

\begin{figure}
   \centering
    \includegraphics{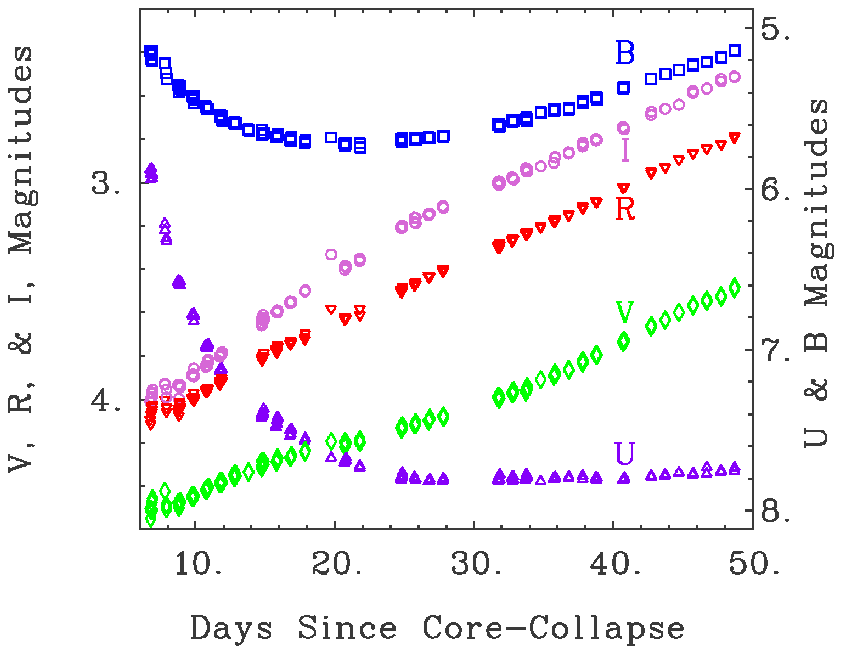}
   \caption{
The U, B, V, R, and I points from the CTIO 0.41-m from
days 6 to 50 (see $\S$\ref{early}).
               }
              \label{ubvri}%
    \end{figure}

This decrement is preceded by a {\it spike} ('E' in Fig.~\ref{fearly},
Fig.~\ref{DEF}, middle, and Fig.~\ref{ubvri}, day 19.8) of up to
10$^{40}$ erg s$^{-1}$ in the CTIO data, with the unusual colors
of B, R, and I, in ascending order, very close to the B and I bands
speculated for the 2.14 ms signature observed from 1987A \citep{M00b,Ha87}
with an H${\alpha}$ enhancement.
Spectra taken by \citep{Ha87} around March 15
(the `Bochum event'), 
\citep{Men87} on March 14.820 (day 19.504),
and \citep{Da87} on UT 1987, March 15.08 (day 19.76), 
support these flux enhancements, including the H${\alpha}$ 
(and a Ca II triplet, ${\rm {\lambda \lambda}}$849.8, 854.2, 866.2 nm, 
as well).  The timing of this event, one
day prior to the decrement, suggests that it is due to a photon stream,
stripped of its UV component by absorption (the CTIO U point at day 19.8
was low, consistent with this interpretation), scattering into other
directions, including the line of sight to the Earth, by what might
have been a clumpy end to the circumstellar material.  Optical pulsations
were not detected at this early epoch (R.~N.~Manchester, private 
communication 2007), because of the oblique view, and the dimensions of 
the beam ($\sim$1 $\ell$t-d).

For the geometry
derived in $\S$\ref{geom} below, the one day delay implies
at least the same maximum jet velocity (0.95 c), supporting this
interpretation of a beam and jet penetrating extrastellar material, 
and giving us a rough isotropic lower limit estimate
of the strength of the pulsations.  Spectra taken just before day 5,
showing an enhancement for wavelengths below 5000 \AA, explain the
discrepancy between the CTIO and FES points at that time
(Fig.~\ref{fearly}).

In spite of the coincidence between the end magnitude of the linear
ramp and that of the Mystery Spot, the two are probably not the same
effect, as the offset of the Mystery Spot from SN 1987A was only 0.045 arc
s even 10 days later at day 30 (Fig.~\ref{spot}), a location barely 
beyond where the
ramp began, as is shown below, and there is no sign of elongation
toward SN 1987A proper in Fig.~1 of [3] or
Fig.~2 of [2].  The Mystery Spot may develop as a plume
within the polar ejecta, pushed by a less collimated, 0.5 c pulsar
wind, perhaps not unlike that observed from the Crab pulsar
\citep{He02}, many days after the passage of the initial, very fast,
very collimated component of the jet.  A beam only 1 $\ell$t-d across
at $\sim$10 $\ell$t-d translates via Eq.~(1) into plasma at $\sim$20 $R_{\rm LC}$.

It is unlikely that the early light curve is due to shallow penetration
of a precessing jet into a varying entry point of the polar ejecta,
because the high density required to limit jet penetration
comes with a higher opacity which would make the linear ramp hard
to produce in this, the inner boundary of the approaching axial feature.
The requirement for a 0.5 c mean motion of the Mystery Spot between
days 30 and 38, slowing to 0.35 c between days 38 and 50 (ostensibly due
to swept up matter), would also be difficult to fulfill in these
circumstances.  On the other 
hand, the polar ejecta density can not be so low as to allow more
than 1 $\ell$t-d penetration by the enhanced UV flash, or the
drop in luminosity from day 7.8 to day 8.8 would not be as big.  If the
{\it jet} penetration is deep, precession and/or changes in the plasma density
beyond the pulsar light cylinder (Eq.~1), may make its initial track,
within the polar ejecta, a helical cone, and this may assist 
the $\sim$0.5 c wind in the creation of the plume which forms the 
Mystery Spot within three weeks of the initial jet penetration.

We will assume that the optical flux from the interaction between
jet particles and the polar ejecta will not be significantly occulted
in the ejecta itself in the direction to the Earth, otherwise again,
the linear ramp would be difficult to produce.  As we will find
below that the axis of the SN 1987A bipolarity is $\sim$30$^{\circ}$
from the normal to the ring planes, the reason remaining a mystery
even today, this is not necessarily a given.  Proceeding nevertheless:
by scaling homologously inward a factor of 10 from the equatorial ring
density of 10$^4$ cm$^{-3}$, we arrive at a polar ejecta density
estimate of $\sim$10$^7$ cm$^{-3}$ -- sufficient to stop the UV flash
from penetrating $>$1 $\ell$t-d.

Assuming a polar ejecta depth, D, of $\sim$10
$\ell$t-d, or 2.6$\times$10$^{16}$ cm, gives a total column of
2.6$\times$10$^{23}$ cm$^{-2}$, sufficient to hopelessly disperse
any pulsed radio signature, though enough to warrant some concern
about continued particle penetration.
However, only a fraction of the protons in the jet will
scatter through the entire depth of the polar ejecta (the slight
concave downward departure from linearity most apparent in
the CTIO data, between days 9 and 20, may reflect this loss,
and/or the density in the polar ejecta may decrease with distance).
In addition, we will find that the angle, $\theta$, from our line of
sight to the SN 1987A beam/jet, will be large in the self-consistent
solution, justifying our assumption of visibility for
the luminous column within the polar ejecta between days 9 and 20.

\section{The Geometry and Kinematics of the Beam/Jet}
\label{geom}

Using the constraints shown in Figs.~\ref{magdist} and \ref{fearly},
we can solve for the
three geometric variables, $d$, $D$, and $\theta$, diagrammed in
Fig.~\ref{fgeom},
and the maximum velocity of the particles in the jet, $\beta$.
First the UV flash hits the beginning of the polar
ejecta at day 7.8:
\begin{equation}
d(1-cos \theta)  = {\rm c}t_0 == 7.8~{\rm \ell t-d},
\end{equation}
where $d$ is the distance to the beginning of the polar ejecta,
$\theta$ is the angle from the beam/jet/polar ejecta direction 
to the line of sight to the Earth, and c is the speed of light.

From Fig.~\ref{fearly} we also have the jet particles well into
the polar ejecta by day 9.8.  Extrapolating the three CTIO points
(just after day 8) backward per above, we have the fastest beam 
particles hitting the polar ejecta at day 8.26:
\begin{equation}
d(1/\beta-cos \theta ) = {\rm c}t_1 == 8.26~{\rm \ell t-d}.
\end{equation}
Next, we have the projected offset of 0.060 arc s for the Mystery Spot,
measured at day 29.8 by \citep{Ni87} and refined
by \citep{Ni99}.  This is more difficult to pin down relative to 
its position radially through the polar ejecta,
so we assume it's some fraction, $\alpha$, of the way through
the polar ejecta depth, $D$, and derive a self-consistent solution:
\begin{equation}
(d+\alpha D) \sin \theta = {\rm c}t_2 == 17.3~{\rm \ell t-d},
\end{equation}
using 50 kpc for the distance to SN 1987A.
Finally, we have the decrement in the light curve at day 20,
shown in Fig.~\ref{fearly}, which we will interpret as the fastest
`substantial' bunch of particles in the jet breaking
through the end of the polar ejecta:
\begin{equation}
(d + D)(1/\beta - cos \theta ) = {\rm c}t_3 == 20~{\rm \ell t-d}.
\end{equation}

The solution to Eqs.~(4-7) gives a constant ratio for $D$ to $d$,
independent of $\alpha$:
\begin{equation}
d  = D t_1/(t_3 - t_1),~{\rm or}~(d + D) = d~t_3 / t_1,
\end{equation}
while $\theta$ is given by:
\begin{equation}
\theta  = 2 \tan^{-1} \{{t_0 \over t_2}(\alpha({t_3 \over t_1}-1)+1)\}.
\end{equation}
The parameters, $d$, $D+d$, and $\theta$ are plotted against
$\beta$ in Fig.~\ref{soln} for 0.3 $\le$ $\alpha$ $\le$ 0.7,
along with the maximum $d$ and minimum $D + d$ implied by the
three measurements of the
Mystery Spot angular separation at days 30, 38, and 50.
Figure \ref{soln} shows that the polar ejecta, at the very least,
must start by 14 $\ell$t-d or closer, and extend to 22 $\ell$t-d or farther,
consistent with our early 10 $\ell$t-d estimates for $d$ and $D$,
as is the high value of $\theta$ ($65^{\circ}<\theta<85^{\circ}$),
which also means that the axis of bipolarity is $\sim$30$^{\circ}$
from the normal to the ring planes \citep{Su05}.

   \begin{figure}
   \centering
   \includegraphics{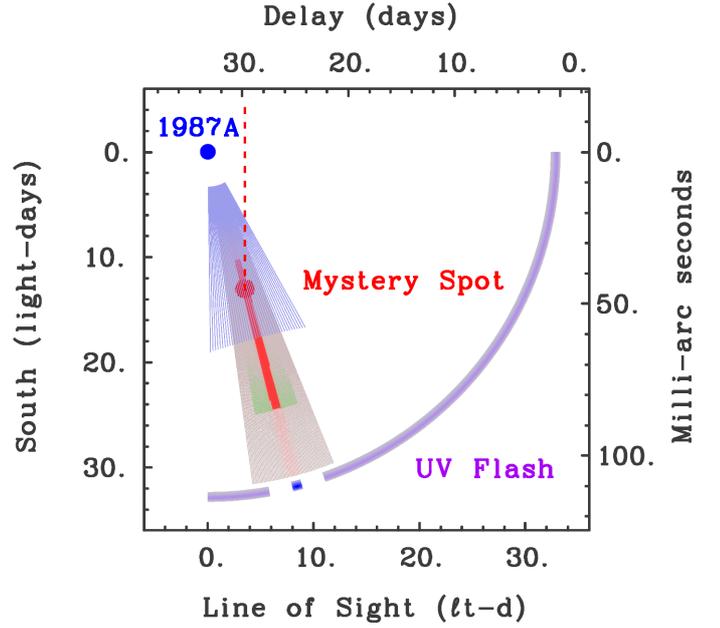}
   \caption{ 
   The relation of the Mystery Spot, near day 30, to a jet,
   a thinning shroud, and a UV Flash, when its projected offset 
   from SN 1987A was 0.045 arc s.
               }
              \label{spot}%
    \end{figure}

   \begin{figure}
   \centering
    \includegraphics{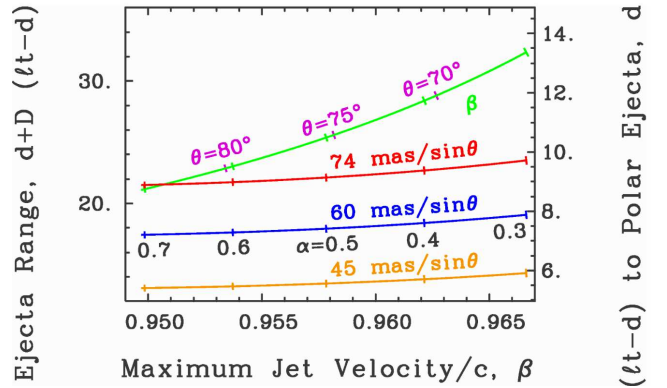}
   \caption{
   The solution values for Eqs.~(4-7). (Horizontal)
   The maximum jet velocity, $\beta$.  (Left vertical) The maximum
   range of the polar ejecta.  (Right vertical axis) The distance 
   from the pulsar to the beginning of the polar ejecta.  
   The line with the steepest slope
   converts $\beta$ (bottom) to $D + d$ (left), or $d$ (right), and
   three values for $\theta$ are marked.  The three other lines
   with moderate slopes constrain the minimum of $D + d$ (right
   end of 74 mas curve), and the maximum of $d$ (left end of the
   45 mas curve and also read on the left vertical axis) from offset
   measurements of the Mystery Spot, which is assumed to be
   a jet-driven plume {\it within} the polar ejecta.  Five values
   of $\alpha$, from 0.3 to 0.7, are ticmarked on the four curves.
               }
              \label{soln}%
    \end{figure}

Given the similar magnitudes of the early lightcurve and the Mystery 
Spot (and indeed, the two are just phases of the same jet phenomenon,
the first more penetrating and the second more energy-injecting),
the energetics of the jet producing the early light curve and these
are the same as that posited for the Mystery Spot in [1],
except that the early lightcurve phase is shorter.  For an
interval of 10$^6$ s, at 5$\times$10$^{40}$ erg s$^{-1}$, the optical
output, from reprocessing of X-rays resulting from the jet particles 
scattering with electrons, which then add to it through
free-free or later synchrotron radiation, is 5$\times$10$^{46}$
erg.  Since only a fraction of the particles scatter
in the polar ejecta, the overall efficiency, in the conversion of kinetic
energy into optical luminosity, could still be as low as
Meikle et al.'s estimated 0.001, which gives 5$\times$10$^{49}$
erg of kinetic energy in the initial jet.  For 0.9 c protons,
each with 0.002 erg of kinetic energy, this would mean 2.5$\times$10$^{52}$
protons or 2$\times$10$^{-5}$ M$_{\bigodot}$ initially each jet.
Without the now visible counterjet, the `kick' velocity to
the neutron star would only be 10 km s$^{-1}$, thus such jets
are unlikely to produce the 100s of km s$^{-1}$ velocities seen
in a few other pulsars.

For a pulsar with
an initial spin rate of 500 Hz this short phase alone would
result in a drop of 10 Hz, corresponding to a mean spindown
rate of 10$^{-5}$ Hz s$^{-1}$, assuming a neutron star moment of
inertia of 5$\times$10$^{44}$ g-cm$^2$.
This may still be an underestimate, as accelerating a square 
$\ell$t-d of the polar ejecta column, which amounts to 
$\sim$0.002 M$_{\bigodot}$,
to $\sim$0.3 c, requires 1.6$\times$10$^{50}$ erg of kinetic energy, 
which amounts to 6.6\% of the 2.5$\times$10$^{51}$ erg of rotational 
energy of a 500 Hz pulsar for each jet, or $\sim$66 Hz of frequency 
drop from 500 Hz, still assuming 100\% conversion of jet kinetic energy
into Mystery Spot kinetic energy, unless the plume has a smaller cross 
section than 1 $\ell$t-d$^2$, and/or the polar ejecta is less dense, on
average, than 10$^7$ cm$^{-3}$.  Since these numbers account only for the
near polar SLIP beam(s), and not to the more equatorial beams, they must
be considered to be lower limits.  Observations of initial pulsar 
spindowns (see $\S$\ref{link}) would help greatly in resolving this
uncertainty.  Spinup from accretion may temper
the spindown somewhat \citep{Pat09}, but gravitational
radiation reaction may counter it \citep{Owen98}, though
the high electromagnetic spindown will mask any effect
of this latter on the pulsar braking index, $n$, where
${\partial f \over \partial t} \propto -f^{n}$, and
$n=5$ for pure gravitational radiation.

In either case, the rotational energy required is too large
to be supplied by a strongly magnetized pulsar over the required 
timescale, unless these are born spinning faster than the moderate
rates generally believed to be typical (e.g., the $\sim$12 ms initial
period for the 16.1 ms PSR J0537-6910 \citep{M06} --
see $\S$\ref{0537}).  
There was certainly no evidence for a strongly magnetized pulsar 
within SN 1987A in its first few years (e.g., [38-40])
and most importantly, 
there is no evidence for such a pulsar in the last few years \citep{Gr05}, 
whereas SN 1986J, at the same age, showed clear evidence of such a
pulsar within it.\footnote{This SN, in the edge-on spiral
galaxy, NGC 0891, exceeds the luminosity of the Crab nebula at
15 GHz by a factor of 200 \citep{Bie04}, and thus
is thought to have occurred because of a core collapse due to
iron photodissociation catastrophe 
(FeSN), producing a {\it strongly} magnetized neutron star ($\sim$10$^{12}$
G -- the origin of magnetic fields in neutron stars is still poorly
understood, though it is believed that thermonuclear combustion
in a massive progenitor to an Fe core is related).}

However, there
may be a weakly magnetized pulsar within SN 1987A \citep{M00a,M00b}, and
at the very least this is supported by solid evidence for the
at the very least this is supported by solid evidence for the
may be a weakly magnetized pulsar within SN 1987A \citep{M00a,M00b}, and
at the very least this is supported by solid evidence for the
formation of a neutron star \citep{Bi87,Hi87}.
A binary merger of two electron-degenerate stellar cores
(DD -- in isolation these would be white dwarfs) has been
proposed for SN 1987A \citep{Pod89}, and the triple ring structure,
particularly that of the outer rings, has recently been successfully
calculated within this framework \citep{MP07}.
Other details of
SN 1987A, including the mixing \citep{Fr89}, the blue
supergiant progenitor \citep{Sk69}, the early polarization 
[49-51], 
and the possible 2.14 ms optical
pulsations \citep{M00a,M00b}, support this hypothesis.

The first clear evidence for DD-formed millisecond pulsars coincidentally
came in the birth year of SN 1987A, with the discovery of the 3 ms pulsar,
B1821-24 \citep{Ly87}, in the non-core-collapsed globular cluster,
M28.  Subsequently many more were found over the next 20 years in such
clusters (e.g., 47 Tuc -- \citep{Ca00}), and attributing these to
recycling through
X-ray binaries has never really worked 
\citep{CMR93,GB10}, by a few orders of
magnitude.\footnote{Attempts to get around the discrepancy between
the cluster ms pulsar population and their very much less numerous
X-ray binary supposed progenitors, invoke a short time interval 
with a very high spinup 
rate, ostensibly produced by a very high accretion rate in the X-ray 
binaries, to generate their ms pulsars.  There are
two problems with this hypothesis.  The first is that
relatively slowly rotating, recycled pulsars weighing
1.7 M$_{\bigodot}$, in the core-collapsed globular cluster, Ter 5 
\citep{Ra05}, have removed high
accretion rate from contention as an alternative mechanism to produce the
ms pulsars in the non-core-collapsed globular clusters without a long
duration (and more obvious) X-ray source phase (see below
regarding J1903+0327).  The second problem is that we will see
in $\S$\ref{LMXBs} that because of SLIP the relation
between accretion rate and spinup rate is often inverse.  The three
ms pulsars in Ter 5 O, P, and ad \citep{He06}, and five other 
non-GC pulsars
with periods $<$ 2 ms, may have been recycled {\it starting} with periods 
near 2 ms.  
}
Thus the DD process in SN 1987A, albeit within a common envelope, would
likely have formed a rapidly spinning, weakly magnetized pulsar.

Consequently, we also argue, as a corollary implication useful for
understanding the SN process and its modern-day observation history, 
that the vast majority of core-collapse events are similar to SN 1987A, 
in that they are a result of the DD process, producing 
only weakly magnetized, rapidly-spinning, ms pulsars, the notable
exceptions being SNe 1986J, and 2006gy and 2007bi, these latter which 
will be discussed further below.  In spite of the rarity of recent
SNe with solid evidence for a strongly magnetized neutron star
remnant, many still currently hold that such a result is the more
frequent result of stellar core-collapse events.

   \begin{figure}
   \centering
   \includegraphics{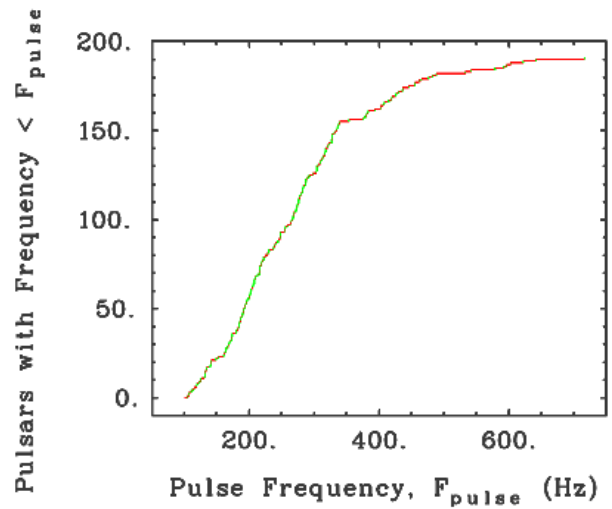}
   \caption{
   The distribution of pulsars from the ATNF pulsar database 
   with spin rates greater than 100 Hz
   (http://www.atnf.csiro.au; \citep{Man05}).
   }  
              \label{PopFreqs}%
    \end{figure}

The above estimates for an early phase of drastic spindown are in
line with the differences between the years-old spin period of the
2.14 ms (467 Hz) signature for SN 1987A of \citep{M00a,M00b} as a
Population I pulsar,\footnote{The 2.15 ms pulsar, J1903+0327 \citep{Ch08}, 
has a main sequence companion in a highly eccentric ($\epsilon = 0.44$), 
95-day, $>$105-$\ell$t-s orbit, from which it could never have accreted mass, 
but its high mass of 1.67 M$_{\bigodot}$ \citep{Fr10} indicates an extended 
era of accretion, ostensibly from a closer companion which has since
been evaporated by the pulsar.  Whether this accretion converted a strongly 
magnetized, only moderately rapidly spinning neutron star into J1903, at 
the same time drastically attenuating its magnetic field, rather than just
converting a light J1903 into a heavier one, is still an unresolved
question, given the effectiveness of weakly magnetized, rapidly spinning 
pulsars in driving jets, where jet-driven spindown might easily dominate
accretion-driven spinup. Thus J1903 might have been recycled from its
lighter self to its heavier self with little net change in spin rate
(see $\S$\ref{Recycling}).} 
and the Population II pulsars with spin periods just over 2 ms, such as
Ter 5 Y and V (2.05 \& 2.07 ms -- 488 Hz \& 482.5 Hz), and 47 Tuc F 
(2.10 ms -- 476 Hz).  The pulsar, J0034-0534, has the next {\it highest}
spin rate, at 532.7 Hz, a gap of 44.5 Hz.  There are {\it seven} in
the same frequency range {\it below} Ter 5 Y (not counting SN 1987A), lending 
credibility to the commonality of 2 ms for the birth periods for pulsars 
produced by DD
merger.\footnote{Because there is initially always too much angular 
momentum, this rate is set by the branching of the Maclaurin and Jacoby 
solutions, leaving a near-Chandrasekhar mass white dwarf spinning with
a period of 1.98 s.}
This would
also explain how ms pulsars could be recycled to spin rates faster than
500 Hz \citep{GL78}, without requiring any decay of the magnetic field.
The occurrence of pulsars with various spin frequencies shows an increased 
density for frequencies just under 500 Hz, consistent with this picture (see
Fig.~\ref{PopFreqs}).  After a (significant) gap from 378 to 346 Hz, where
there is only PSR J0024-7204S at 353.3 Hz, the pulsars slower than 346 Hz, 
the steepest part of the curve, comprise the `5 ms pileup,' i.e., those
pulsars (in part) possibly the offspring of defunct LMXBs, are discussed 
further in $\S$\ref{Recycling}.  The pulsars between 378 and 500 Hz
are not numerous enough to populate the pileup just by spinning down,
without resorting to entities such as age and beaming differences.

\begin{table}[!h]
\caption{15 M$_{\bigodot}$ DD-formed pulsar processes}    
\label{table:2}     
\begin{tabular}{l l l }       
\hline\hline                 
Process   &&Timespan \\   
          &&t$_{min}<$t$<$~t$_{max}$ \\    
\hline                        
 Bi-lobed to spheroidal CE                       &&$\sim$-2,000? years \\
 Chandrasekhar-mass WD formed                    &&$\sim$-2? seconds \\
 DD core-collapse to 2 ms pulsar                 &&$\sim$-0.5$<$t$<$0.5? s \\
 Gamma-ray burst ($\ell$GRB)                     &&$\sim$~0.5$<$t$<$$\sim$100? s \\
 Pulsed GRB afterglow                            &&$\sim$~0.5 s$<$t$<$1 day \\
 Ejection spindown phase, $\theta_{\rm V} \ll 1$ &&~~~~0$<$t$<$2 years  \\
 Transition to $\theta_{\rm V} \sim 1$           &&~~~~2$<$t$<$4 years  \\
 GR-dominated spindown                           &&~~~~3$<$t$<$100? years  \\
 Broadly beamed pulsations                       &&~~~~4$<$t$<$10 years  \\ 
 Plasma depleted                                 &&~~10$<$t$<$$\infty$ years  \\ 
 Cas A-type thermal X-ray source                 &&~~10$<$t$<$400 years  \\ 
 Magnetic dipole-dominated spindown              &&100$<$t$<$$\infty$ years  \\
\hline                                   
\end{tabular}
\end{table}

Table \ref{table:2} lists the anticipated emission and spindown phases 
of a SN/pulsar similar to 1987A.  The epoch at which the common
envelope transitions from bilobed to spheroidal is estimated 
at 2000 years prior to core-collapse, at which time the spindown
is dominated by SLIP losses to mass ejection/disruption of the 
progenitor star, and thus very high.
The time range at which the losses transition from SLIP to
gravitational radiation is estimated to be 2 to 4 years, after
which GR losses are estimated to dominate for a century.  
At -2$\times$10$^{-10}$ Hz s$^{-1}$, this amounts to a loss of 
$\sim$0.6 Hz, which may be an overestimate, since the spindown 
may not sustain this rate over a century.  Certainly the loss can 
not amount to several dozens of Hz, as the resulting tight grouping
of pulsars with spin periods near 2 ms would not happen.
When the pulsar is about a decade old, there is not enough
plasma left (barring occasional clumps) to sustain optical
pulsations.  Thus the epoch of optical pulsations for SN
1987A started near 4 years of age, which is relatively late, 
due to the large angle between the spin axis inferred from
the bipolarity, and the line of sight to 
the Earth, some 75$^{o}$.  

In this case, plasma initially available at many
$R_{\rm LC}$ resulted in axially driven pulsations.  Precession and 
nutation may have further embellished the axial pattern \citep{M00a}.  
A strongly magnetized pulsar, because it will be spinning more
slowly than its weakly magnetized counterparts, will have a larger
$R_{\rm LC}$ and thus produce a much less collimated beam and jet 
than those of SN 1987A because plasma 
is available at far fewer $R_{\rm LC}$, but the larger progenitor 
counteracts this trend somewhat.  As the 1987A circum-neutron star 
density declined, polarization currents were restricted to fewer 
$R_{\rm LC}$, resulting 
in pulsed emission along a cone of finite polar angle, which
may have modified the resulting jet into the approaching 
and retreating conical features now easily visible in the 
HST ACS images \citep{Wa02}.  The high velocity cloud, HVC A0
\citep{VN11}, may be an example of a pulsar-driven jet long
separated from its progenitor.

Eventually, as the plasma continued to thin with time, its maximum 
density just outside of $R_{\rm LC}$, occurred between here and 
2$R_{\rm LC}$, resulting, because of Eq.~(1), in pulsations driven 
close to the pulsar's rotational equator, and according to our 
self-consistent solution, in the line of sight to the Earth.  
Such a beam can not possibly produce 
the observable excess luminosity that may
have been seen by 1991 [63-65], as
the amount of $^{57}$Co required to otherwise account for
the excess (m$_V \sim$17.4 $\sim$275 L$_{\bigodot}$) was only barely consistent with hard X-ray
and infrared spectral data \citep{Rank88,Kum89}.  
If we were in the cusp of the pulsar within SN 1987A during this
epoch, because of the distance$^{-1}$ law associated with this
direction, the implied luminosity would have dropped from
10 L$_{\bigodot}$ (roughly magnitude 21), by half the distance 
modulus to the LMC, some 9.4 magnitudes, or a factor of 
$\sim$5,000, to $\sim$0.002 L$_{\bigodot}$.  Moving
from 10 pc to $\sim$12 $\ell$t-d (0.01 pc), the likely distance 
to the equatorial band (velocity $\sim$2,500 km s$^{-1}$), the 
apparent luminosity falls by another factor of 1,000,
far short of the required excess, and all possible other beams
can not possibly make any difference.  Because our 
geometrical solution puts the pulsar behind the equatorial
torus visible in ACS images taken years later, the excess
luminosity may have been due to an accretion-powered X-ray flux,
reprocessed within ejecta thinning with time.

A few
years earlier it is unlikely that the 2.14 ms signal would have
been detectable in the broadband, as limits established in early
1988 \citep{Pe89} are comparable to levels of the 2.14 ms signal
observed in the I band between 1992, Feb.~and 1993, Feb.  The
2.14 ms pulsar candidate was last detected in 1996,
Feb.~\citep{M00b}, and by 2002 there was no evidence of
any such source in ACS images, which only really means 
that any pulsar within SN 1987A had entered the
`Cas A' phase,\footnote{Calculations with the SLIP model
involve a variation of Kepler's equation, which relates the
eccentric anomaly, $E$, to the mean anomaly, $M$, using the
eccentricity, $\epsilon$, $\rm{E} - \epsilon \sin \rm{E} = \rm{M}$,
but in this case $\epsilon > 1$.  Such calculations are notoriously
difficult, even for a compact star {\it not} surrounded
by remnant plasma.  Needless to say, no such calculations
have been done to date, and thus no calculation through the
core-collapse process, done so far, possibly including those of 
collapsars, can possibly be valid.  One side effect of not 
properly accounting for the pulsar, and the large amount of 
$^{56}$Ni, when strongly-magnetized pulsars are produced, is a 
very low estimate for the mass, $\sim$25 M$_{\bigodot}$, above 
which the core collapse continues 
on to a black hole.  SNe 2006gy and 2007bi, each with several 
M$_{\bigodot}$ of $^{56}$Ni, argue for a much higher 
transition mass.}
having exhausted its surrounding excess plasma
and perhaps also because the Earth was no longer
in the `cusp' of its beam(s).  Still, the central
source should turn on when the pulsar encounters matter
from time to time.  Thus continued observations are warranted.

\section{The SN 1987A Link to GRBs}
\label{link}

Without (or perhaps even with)
the H and He in the envelope of the progenitor of 1987A
Sk -69$^{\circ}$202, the collision of the jet with the 1987A polar 
ejecta (which produced the early light curve and Mystery Spot)
might be indistinguishable from a full $\ell$GRB 
\citep{Cen99}, otherwise it would just beg the question
of what distant, on-axis such objects {\it would} look like.
It is also the case that few or no $\ell$GRBs have been found in 
elliptical galaxies, and that the DD process {\it must} dominate (as
always, through binary-binary collisions), by a large factor the 
neutron star-neutron star mergers in these populations, even when 
requiring enough white dwarf-white dwarf merged mass to produce 
core-collapse.  All of these facts/realizations lead to the 
alternate conclusion that the DD process may produce sGRBs in the absence 
of common envelope and polar ejecta, the means by which they would 
otherwise become $\ell$GRBs.  If the sGRBs in ellipticals 
are due to mergers of white dwarfs, we can conclude that:
1) the pre-common envelope/polar ejecta impact photon spectrum of
$\ell$GRBs may be well characterized as those of sGRBs,
2) sGRBs are offset from the centers of their elliptical hosts 
possibly because
they are white dwarf-white dwarf mergers to core-collapse in their hosts' 
globular clusters (to produce most of their ms pulsars -- 
\citep{Gh5}), and 3) neutron star-neutron star mergers may not 
make GRBs as we know them, and/or be as common as previously thought.  Thus
sGRBs, which last a few seconds \citep{Ha01}, may not flag neutron star-neutron 
star mergers, which may last only a few ms, the same timescale as the 30-Jy, 
DM=375 radio burst \citep{Lo07}. 

Supernova 1987A, with its beam and jet producing its early light 
curve and Mystery Spot, is therefore potentially the Rosetta Stone 
for three of the four types of GRBs: $\ell$, i, and s
GRBs \citep{Hor06,Hor09},\footnote{All except Soft Gamma Repeator
(SGR) GRBs, which are estimated to amount to less than 5\% of sGRBs
and 1.5\% of the total \citep{Pa05}.}
with both polar ejecta and common envelope, red supergiant common
envelope and no polar ejecta, and neither polar ejecta nor common
envelope, respectively \citep{M07}.

\begin{figure}
   \centering
   \includegraphics{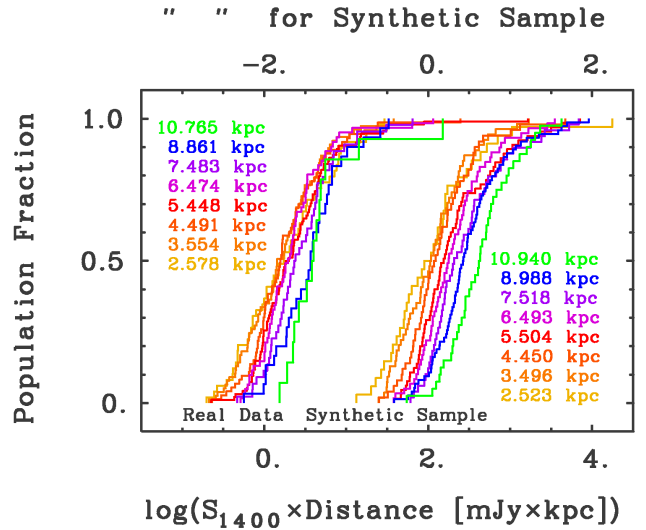}
   \caption{
   (Left, lower horizontal axis) The population of 497 pulsars from the 
   Parkes Multibeam Survey which fall into binned distances, with centers 
   between $\sim$2.5 and $\sim$10.75 kpc, plotted against the log of the 
   product of their 1400 MHz fluxes and their NE2001 distances from 
   \citep{Co02,Co03}. (Right, upper horizontal axis) The same for 
   a synthetic distribution of pulsars which obey the inverse square law.
   The two families of curves are both in order of increasing distance
   from left to right at a population fraction of 0.2.
    }  
              \label{logslogd}%
    \end{figure}

In addition to axially driven pulsations, the SLIP model makes the
very remarkable prediction that the component of pulsar
intensity which obeys Eq.~(1), diminishes only as distance$^{-1}$
(see the Appendix),
and verified experimentally 
\citep{Ar04},\footnote{Anyone 
who has noticed the wildly varying acoustic signature from helicopters
has experienced this directly \citep{Lo96,My89} and similar effects in
other geometries are also well known in acoustics \citep{Go52,Li53}.}
and has been dramatically confirmed \citep{SSM09} and noted by
others \citep{Lo11}, for pulsars in the 
Parkes Multibeam Survey (e.g., \citep{LFL06}) with single, 
narrow pulses (see Fig.~\ref{logslogd} and
$\S$\ref{TypeII}).\footnote{The pulsars at high flux 
in Figure \ref{logslogd} neatly overlap when multiplied
by their distances, confirming the distance$^{-1}$ law,
while those with low fluxes are detector-limited and thus
are ordered according to their distance.  The synthetic
sample was generated with a distance$^{-2}$ law, and
are by contrast spread out according to their distance
even for the parts with the highest fluxes/population
fractions, well in excess of their detector-limited parts 
near the bottom at low fluxes.

The subbeams 
which make up the cusp radiation have a constant height parallel
to the rotation axis ($\delta(\theta_{\rm V})\times {\rm distance} = 
{\rm constant}$) 
and thus, to conserve energy their flux drops only as distance$^{-1}$.  
However, because they do have a constant height (of order of several
$R>R_{\rm LC}$), the solid angle they subtend also decreases 
as distance$^{-1}$, and thus energy is again conserved \citep{Ar07}.  
However, as long as pulsar survey sensitivities keep up with the 
distance$^{-1}$ law, and their survey beams are always filled with 
productive target, such as a galaxy, they will always detect {\it 
more} new pulsars with distance. }
There is also evidence that GRBs and their afterglows share 
this characteristic \citep{KK10,KK11}, which supports the SLIP prediction 
of axially driven pulsations when plasma is available at many 
$R_{\rm LC}$.  The SLIP prediction also explains how GRBs and their
afterglows can be visible across the Universe.
As a consequence of this prediction, we initiated a campaign
of high speed monitoring of GRB afterglows in 2008 and 2009
at the Lick Observatory Crossley 36-inch telescope
(without success -- `live' bright afterglows occur about
one night in 60), and will re-initiate it as soon as the 
opportunity presents itself.

If, as for SN 1987A, the vast majority 
of SNe are DD-initiated, then by measuring the pulse period,
$P$, of the optical/near infrared pulsations from an afterglow,
and assuming the pulsars resulting from DD are all produced
at a standard spin period, $P_0$, first measured from SN 1987A
near 2.14 ms, but corrected to near 2.00 ms because of material
ejected, the redshift is given by:
\begin{equation}
z = {P \over P_0} - 1,
\end{equation}
and even a moderately precise $P$ (by standards), may yield a
very precise redshift.

   \begin{figure}
   \centering
   \includegraphics{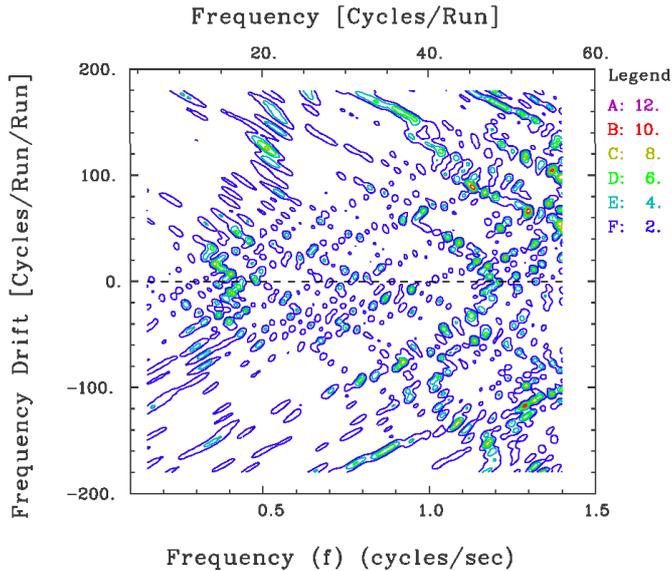}
   \caption{
   Low frequency Fourier power from GRB 960216 (with $\sim$18,000
   events in 40 s) is contoured on the frequency-frequency derivative 
   plane.}
              \label{f-fdot}%
    \end{figure}

SLIP predicts that the peak of the emission for slowest pulsars occurs
in the gamma-ray band [85-87], and this is supported by recent
gamma-ray detections of slow ($\sim$1 Hz) pulsars in supernova remnants
by FERMI (e.g., \citep{Ab08}).  Without this feature, one would naively
question whether SLIP could produce pulsed GRB afterglows when the predicted
post-core-collapse strength of the magnetic field at 100s of $R_{\rm LC}$, whether
from a several Teragauss 80 Hz pulsar, or a few Gigagauss 500 Hz pulsar, would
be well below the few megagauss necessary to drive cyclotron radiation in
the optical and near-infrared bands.  SLIP is more complicated: the
centripetal acceleration of the polarization currents leads to contributions
for a wide range of high frequencies.  The essential necessary ingredient is 
plasma (see $\S$\ref{Plasma}), and in this situation, there's plenty.

There is no requirement in the SLIP
model on the rotator being a neutron star -- a white dwarf 
will do as long as it has a magnetic field and there is
plasma outside of its light cylinder, whether this comes from
pulsar-generated plasma sheets, or gravitationally attracted ISM
or SNR plasma.  If this is the case,
strongly magnetized pulsars may not make GRBs, 
and it might even be possible for a pre-core-collapse,
$\sim$1.4 M$_{\bigodot}$ white dwarf, spinning at its minimum period of
1.98 s, to emit the prompt part of a GRB, and, as with the afterglow,
the distance$^{-1}$ law would likewise ameliorate the energy
requirement, even with the large expected spinup.  This also 
raises the intriguing possibility that a GRB could be produced
without core-collapse, and a large spin-{\it down} may be present.


We tested for spinup/down in the GRB with the highest fluence in
the BATSE catalog, 960216 \citep{Pa99}, by Fourier transforming the first
40 s of events and contouring power on the frequency-${\partial f \over
\partial t}$ plane.  Power appears, though not significant
without further confirmation, at a mean frequency of 0.50 Hz, and
derivative of +0.08 Hz s$^{-1}$, and also for spinup/down
about an order of magnitude smaller, in the 0.35 to 0.42 Hz region
(the results for higher frequencies, assuming 0 $\dot{{\rm f}}$, are also
negative).
Bursts with even better statistics (perhaps from FERMI)
may be necessary to further test this hypothesis.

It is difficult for a GRB produced by pulsations during spinup to 
produce an inverted distance law because of the changing $v$.
Alternately, GRBs may be produced when gamma-ray photons from
the initial, collimated SLIP photon flux scatters off of 
circumstellar material such as polar ejecta \citep{Ar10}.  
For material such as polar ejecta, in a plane  perpendicular to a 
line of sight from the SN proper, the pattern of illuminated, scattering,
concentric rings expands supraluminally, and thus this emission also
exhibits a distance$^{-1}$ law for certain directions.  However,
the initial gamma-rays from the compact remnant actually penetrate
the polar ejecta considerably and thus the scattering produces a
volume source.  What will be observed as a GRB across the Universe
will be whatever produces a supraluminal source with an inverted
distance law.

The geometric model with small angle scattering of gamma-rays in
distant polar ejecta can explain other details of $\ell$GRBs,
such as their $\sim$100 s T$_{90}$'s,\footnote{An offset of
0.5$^{\circ}$ at 10 $\ell$t-d corresponds to a 33 s delay.}
the negligible spectral lag for late ($\sim$10--100 s) emission
from `spikelike' bursts \citep{NB06}, and why `precursor' and
`delayed' contributions have similar temporal structure
\citep{NP02,M07}.

Moving a 10$^{40}$ erg s$^{-1}$ source from 25 $\ell$t-d to 8 billion
light years (z$\sim$1), assuming a distance$^{-1}$ law, would only 
reduce the apparent magnitude to -15.  Optical flashes of visual 
magnitude 2, which reddened as they faded, were observed on two occasions by
Howard Bond (private communication, 2008).  No known triggers occurred
at the noted times.

X-ray afterglows of GRBs are characterized by a power law index, $\alpha$, 
for the decline of intensity with time, and another, $\beta$, to
characterize the power law decline of the X-ray spectrum.  The temporal
index is discontinuous at sudden changes in slope called `jet breaks'.  
The two indices can be related through the `closure relations' for 
various circumstances affecting the jet, although with limited success 
for Swift GRBs \citep{Ra09}.  In the SLIP model, breaks in
the temporal slope {\it of the optical afterglows}, if they existed,
would be caused by sudden changes in plasma density outside
of the neutron star, through Eq.~(1).  Because the steep decline of
the X-ray flux ($\alpha \sim 3$) matches that of the GRB, which
is unlikely to be pulsed, we do not expect the X-ray afterglow
to be pulsed either, and their jet breaks may be caused by
changes in the scattering of high energy photons due to structure 
in the polar ejecta \citep{Ar10}.  However, in order to be observable
at all, the X-ray afterglows must have an inverted distance
law similar to GRBs resulting from a supraluminal process of
one sort or another.  Whether or not these are pulsed at all
remains to be determined.

SLIP has little difficulty explaining the unusually long duration of GRB 
101225A (T$_{90}>2000$ s) through a coalescence between a neutron star and 
another star \citep{Th11,Ca11}, i.e., a `borrowed SN'.  Initially the 
outer layers of the star will be ejected by SLIP, and this activity will 
become progressively more violent with the inspiral of the neutron star as 
plasma becomes available at increasing $R_{\rm LC}$, though the tilt of the
rotation axis relative to the plane of the inspiral will matter.  It is this 
progression that is unusual for the GRB, and is responsible for the longevity
through gamma-rays scattering in previously ejected material.  
A magnetar or, again, the 
formation of a black hole is unnecessary to make the jets (and highly unlikely 
unless the star is extremely massive).  This is yet another case where high time 
resolution observations are sorely needed.

\section{Double-Degenerate in Type Ia/c SNe}
\label{Ia/c}

Since 2007, Feb., it appeared unavoidable that Type Ia SNe were
also DD-generated, one of the causes being the long list of reasons
why Ia's can not be due to thermonuclear disruption 
[74,96-99].  Now, with two out of two SNe Ia 
(SNR 0509-675, SN 2011fe), leaving no symbiotic companion
\citep{Sch12,Bl12}, 
it is not clear if this ever happens in {\it any} progenitor,
and empty SN remnants
almost always contain a neutron star which is not visible, just
as the one in Cas A is barely visible \citep{Ta99}.  Further, this means
that Ia's and Ic's (these latter have been regarded by many as DD-initiated,
neutron star-producing since the invention of the classification), are both
due to the DD process. Thus these must form a continuous class: Ia's when 
viewed from the merger equator, with lines of Fe; and Ic's when viewed from 
the merger poles, because this view reveals lines of the r-process
elements characteristic of Ic's,\footnote{If sufficient matter exists,
in excess of that lost to core-collapse, to screen the Ia thermonuclear
products -- a rare circumstance in elliptical galaxies, the reason why
Ic's are absent from these. }
part of the reason for the differing spectroscopic classification,
and the high approaching velocities\footnote{These velocities, near
0.4 c, match those of the Mystery Spot.}
(e.g., `hypernovae') due to
the view looking down the the axis of the approaching bipolarity.

In the application of Type Ia SN luminosities for cosmological
purposes, the increase in blue magnitude from the light curve maximum,
$\Delta M_B$ -- essentially an inverse measure of the width of the
light curve in time -- in the first few weeks of SN Ia proper
time, is used to correct the Ia luminosity for the variable amount of
$^{56}$Ni produced \citep{Ph93}.
However, the direct relation, between the $\Delta {\rm M _B}$ of the
width-luminosity relation and the fractional SiII polarization in Ia's,
pointed out by \citep{Wa07}, is more meaningfully interpreted
as an {\it inverse} relation between the SiII polarization and
luminosity (unlike the Fe lines, SiII lines must also exist in the
axial features because they are also observed in Ic's, and their
polarization in Ia's is a result of the more rapid axial
extension when viewed close to the merger equator).

This inverse relation would be expected in Ia's if the luminosity
of the (very extended and productive of polarization) axial features 
were fixed, while the luminosity 
of the (less extended and less productive of polarization) toroidal
component is driven by the amount of {\it encapsulated} $^{56}$Ni 
positron annihilation gamma-rays, which can be very
high.  As with 1987A-like events, it
would again beg the question of ``What {\it else} they could possibly
be?,'' and `delayed detonation' \citep{Kh91}, or 
`gravitationally confined detonation' \citep{Pl04}, may not produce
polarization which would be inversely proportional to luminosity.  And  
unless the view {\it is} very near polar,
this geometry can produce split emission
line(s) on rare occasions, as was seen in SN 2003jd \citep{Maz05b},
and thus again there is no need to invoke exotica, or an entire
population (III) to account for GRBs \citep{Con05,M04}.

Because there is a spectroscopic difference between Ia's and Ic's,
the SLIP-driven polar jets are either deficient in $^{56}$Ni, or
are too diffuse to encapsulate their gamma-rays,
or both.  {\it No} observation of {\it any} recent SN other than SN 1986J,
2006gy, and 2007bi, and including all {\it ever} made of Type Ia SNe, is
inconsistent with the bipolar geometry of 1987A.

All this raises serious concerns about the use of SNe Ia in cosmology,
because many Ia/c's in actively star-forming galaxies belong to the continuous
class, and some of these, and most Ia's in ellipticals, may not
encapsulate a sufficient fraction of their gamma-rays to be bolometric,
especially given the toroidal geometry, lying as much as two
whole magnitudes (or more for nearly naked core collapse) below the 
width-luminosity relation (the faint SNe Ia of
\citep{Be05}).  Thus the {\it log} of the systematics (2+ magnitudes)
can be an order of magnitude bigger than the {\it log} of 1 plus the small
effect (0.25 magnitude) from which dark energy was inferred, {\it in the 
same sense}, an effect which could easily swamp Malmquist bias, the largest 
systematic working in the opposite sense.  

In the SLIP model the pulsar eviscerates its
stellar remnant as long as there is remnant remaining, the SN can not 
`close in' to a spherical configuration until much later when it will 
essentially become a shell source.  This enforces a toroidal geometry of 
ejecta near its rotation equator, which may not conform to the width-luminosity
relation because of unencapsulated positron 
annihilation gamma-rays from the toroid, an assertion supported by the early
detection of hard X-rays and gamma-rays [111-114].  However, 
the local sample was
selected on the basis of conforming to the width-luminosity (W-L) relation,
which is by definition \citep{PE01} the high end of the luminosity
function.  Thus the unavoidable inclusion of faint Ia's in the distant
sample leads to a bias which makes it appear to be anomalously dim,
assuming a distance$^{-2}$ law, which seems appropriate for these
objects.
Supernovae which do not conform to the W-L relation have also been noted 
in 2005, by \citep{Kr05}, for high redshift ESSENCE SNe.  


\section{Sco X-1 \& LMXB's, SS 433, and other details}
\label{Details}
\subsection{DD and SN Luminosities}
\label{TypeII}

The double-degenerate mechanism ensures that many SNe are born
from the core-collapse of a post-merger white dwarf with a rotation 
period near 2 s.
In this circumstance the amount of $^{56}$Ni produced depends on 
how much and what type of material is left in the common envelope 
in excess of the $\sim$1.4 M$_{\bigodot}$ lost to the neutron star 
(and emitted neutrinos), which can exceed this. 
The paltry amounts
of $^{56}$Ni associated with Ib's and at least 90\% of IIs are 
likely the result of dilution of their thermonuclear fuel with He
and/or H due to the DD merger process.\footnote{Helium has
been found where it was not expected in almost all well-studied
Type II SNe.}
Thus SNe 2006gy and 2007bi \citep{Sm07,Nu07} do not have to be 
pair-instability SNe 
[119-122], or the former a collision of two
massive stars \citep{PZ07},
only massive FeSNe of up to 75 M$_{\bigodot}$, which may actually 
have produced 3 and 6 M$_{\bigodot}$ of $^{56}$Ni respectively, 
{\it and} a strongly magnetized neutron star remnant, a
prediction which, in the case of SN 2006gy, can be tested
soon.\footnote{Such amounts
of $^{56}$Ni can lift a lot of stellar material, thus
40 M$_{\bigodot}$ in a SN remnant is no longer a reason 
to invoke `millisecond magnetars' as the dispersal 
mechanism \citep{Th04,JV06}}
Alternately, if the progenitor of SN 2007bi was stripped
of its H and He by some mechanism, it could have been
a DD (Type Ia/c) SN with a high luminosity due to a 
high mass (C and O) in its common envelope.

\subsection{SLIP, plasma, and intermittent pulsations}
\label{Plasma}

The presence of plasma makes a huge difference to rapidly 
rotating, weakly magnetized neutron stars.  Strong transient 
pulsations have occurred during observations of SN remnants or X-ray
binaries which have never been subsequently confirmed, and yet
have no explanation other than as a real, astrophysical signal
(e.g., the Cygnus Loop -- see \citep{Bl75}).  Judging from 
the high fraction
of empty SN remnants, the population of `quiet' neutron
stars must exceed all other `loud' populations combined
\citep{Wea94}.  All Galactic pulsars have the interstellar medium
from which to draw plasma with their intense gravitational
field, but frequently this is not enough, then only when 
such a neutron star encounters a cloud of matter will it become
sufficiently luminous to be detected.  It is for this reason
that observations of SN 1987A should continue, as has already been
suggested in $\S$\ref{geom}, in spite of
conclusions drawn from recent `snapshot' null results 
\citep{Gr05,P12}.

SLIP also makes 
no assumptions about the carrier wave -- a physical process such 
as cyclotron radiation/strong plasma turbulence is necessary for 
the pulsar to radiate at shorter/longer wavelengths, but
centripetal acceleration of the emitting region further modifies 
this spectrum by adding emission to shorter wavelengths. 
Thus, whether or not emission is observed in a band where
SLIP predicts it should exist, as, for example, the few
GeV cutoff of the Fermi pulsars \citep{Ab09b}, if real,
would indicate something about the physical conditions of, 
and/or near, the pulsar.

In the context of the SLIP model, emission from rotation 
powered pulsars is not haphazard -- radiation from the 
known pulsars may very well be detected from Earth because 
we are `in the cusp,' i.e., we are in the part of the 
pulsar's beam that decays inversely only as the first 
power of distance (Eq.~1), which may be one of the 
reasons why repeated attempts to detect pulsations from 
extragalactic SNe have failed (e.g., \citep{MK84} -- 
another reason might be that all except SN 1987A were too distant).  

\subsection{Low Mass X-ray Binary (and other) Rotation Rates}
\label{LMXBs}

Although SLIP makes the detection of pulsations from the luminous LMXBs 
unlikely (see the Introduction), there may still be a few ways to tell 
how fast their neutron star components are spinning.
Because neutron stars are the accreting compact members of most, if not
all LMXB's \citep{Liu01}, thermonuclear (Type I) X-ray bursts are common, 
as are the oscillations which occur during them \citep{LB82,Zand}.  
Some of the oscillation 
frequencies are quite high, with 1122 Hz for XTE J1739-285 topping 
the list \citep{Ka07}.  Many of these frequencies 
are thought to reveal the actual rotation period of the neutron
star.  Many, including the first observed burst oscillations in the 
optical, in 4U1254-69 \citep{Ma80}, show oscillations with only modest
frequencies (in this case, 36.4 Hz),\footnote{This  may have 
happened with 4U1728-24, which showed strong 104 Hz optical pulsations 
for a short interval (B.~A.~Peterson \& J.~E.~Nelson, private communication, 
1977 -- if this is the rotation rate, then the $\sim$111 s periodicity is 
due to some other mechanism, see, e.g., \citep{Sa90}).
This result was never published because of its high implied energetics
and it was never reproduced.  However there is no other good explanation given
the single-photon-counting detector used, and SLIP can explain the energetics
(an image of the source at the same time should also have confirmed this), 
and further development of survey telescopes could also detect such events.}
which also might be the neutron star rotation rates.  These modest spin rates
are consistent with the LMXB rotation rates settling to a rate where spindown 
from SLIP, which may increase more than linearly with mass transfer rate, can 
exceed by several orders of magnitude any spinup due to accretion, which 
is subject to the Eddington limit.  The association with rotation 
periods might be made when these are observed to remain the same (or very 
close) in repeated bursts.  
Pulsations from LMXB rotation frequencies may be observed in those
for which the restrictive beaming ($\theta_{\rm V} \ll 1$)
occasionally relaxes, i.e., for which plasma is restricted to
fewer $R_{\rm LC}$ ($\theta_{\rm V} \cong 1$ -- see $\S$\ref{Recycling}).  

There is also a small group of about a dozen transient accreting millisecond 
X-ray pulsars, most of which also only show oscillations during 
Type I bursts (e.g., \citep{Wi10}).
Many of these have accretion rates that are not persistently high, 
and also have very high spin rates, such as 400 Hz for SAX J1808.4-3658.
In addition, of the three fastest pulsars, two, J1748-2446ad and J1959+2048, 
are currently evaporating their extremely low mass companions, and one, 
B1937+21, already has. 
All this is consistent with an inverse 
relation, at some point, between the rates of mass transfer and 
net rates of spinning up.

\subsection{Sco X-1}
\label{SCOX1}

The dual radio lobes of Sco X-1 \citep{FGB01}, which expand
at $\sim$0.3-0.57c, and to which energy must be transported 
at speeds of 0.95 c or greater, are a nearly exact match to the 
derived kinetics of the Mystery Spot of SN 1987A, and might have 
actually been observed and detected from 1987A in the radio band, 
given a sufficiently large and sensitive radio array in the
Southern Hemisphere, which did 
not exist at the time.  Wright et al.~\citep{Wr75} have
noted that the $<$19-hour orbital period constrains the 
companion to a very faint M dwarf, or a white dwarf stellar 
core.  This would imply that the accretia would be rich
in metals as is the case with SN 1987A, though the 0.95 c 
transport of energy in its jets also argues for some hydrogen, 
or at least spall protons (see $\S$\ref{Positrons}),
which are then accelerated by the usual SN `boost' mechanism of 
heavier elements hitting lighter ones.  If so, then considering 
that the elements of the accretia may not be stratified, as they
may have been at least partially so within the newly-born remnant 
of SN 1987A, it is amazing that this mechanism works at all, and 
this, plus the relative dearth of H, may explain the relative 
weakness of Sco X-1's very fast jet component.

\subsection{Historic remnants and pulse profiles}
\label{HistR}

If we are not `in the cusp,' for any annulus of plasma
beyond the pulsar light cylinder, whose induced radiation
is free to propagate to infinity, spinning neutron stars 
may only appear as thermal sources, such as the one in 
Cas A \citep{Ta99}.  A century older, as is the case with
Tycho 1572, 
and not even
the thermal sources are detectable.  However, the extinction
of A$_V$=3.47,+0.13,-0.29  mag.~to the Kepler 1604 remnant \citep{DG80}, 
though a revision up from 2.2 \citep{vdb73}, is still much 
less than the 6.2$\pm$0.6 mag.~to Cas A \citep{Ek09}, with
the distances nearly equal at 3.2 and 3.4 kpc.  Thus assuming
that extinction traces H column and all SNe produce compact 
remnants, the pulsar in this remnant should be visible as a 
thermal source, so perhaps the neutron star in Cas A is 
visible because of SLIP beaming.  Evidence does
linger, however, even in the outwardly very spherical
remnant of SN 1006, as bipolar high energy emission in XMM
and VLA images \citep{Ro04}, the high velocity chimneys seen
in Cas A \citep{MF10} being long gone.  For many pulsars, such 
as the Crab, which were discovered because of a SN remnant association, 
we may not be in the `in the cusp,' and thus the isotropic
optical luminosity of the pulsar may currently still be
$\sim$3 L$_{\bigodot}$.  However, for those pulsars that are 
in the cusp, the SLIP model predicts that the pulse profile
will consist of two very close peaks,\footnote{These are
so close that the singly-peaked profiles of the vast bulk
of radio pulsars could easily be such doubles without
our knowing it (see the Appendix).}
which would seem to exclude the Crab 
(e.g., \citep{Go00} and \citep{Ge11})
and ascribing the interpulse to some other mechanism -- 
\citep{HE07} 
-- see the Appendix), but the solitary pulse of the 
16.1 ms pulsar, J0537-6910, will split, if allowed by the
fitting process (see immediately below).  


\subsection{The pulsar, J0537-6910, and the SLIP model}
\label{0537}

SLIP predicts that pulsed emission arises outside 
of the light cylinder, either from current sheets, originating 
close to the pulsar, or from ISM plasma which has been concentrated by 
gravity from the neutron star, or both.  In either case, the plasma outside
of the light cylinder is most dense {\it just} outside of the light
cylinder.  In the case of the 62 Hz X-ray pulsar in the Large Magellanic 
Cloud, J0537-6910, SLIP explains why it is {\it not} a strong optical pulsar 
\citep{Mi05}.  Although J0537 only has an effective dipole field of $\le$1 
TG, since it is an oblique rotator \citep{M06} its actual surface field 
could be as strong as that of the Crab or B0540-693, 3-5 TG, or even 
stronger.  In addition, because of its rapid, 62-Hz  spin, the magnetic 
field just outside of its light cylinder, which is less than half as
distant as that of the Crab pulsar, is an order of magnitude larger 
still.
This means that the cyclotron frequencies just outside of the
light cylinder of J0537 are an order of magnitude (or more) higher
than those which contribute to the optical pulsations of the
Crab, and thus are shifted out of the optical regime upward
toward the X-ray regime.  Since cyclotron frequencies do not
have subharmonics, J0537 can not, and does not produce strong optical 
pulsations -- the centripetal acceleration of the polarization currents is 
not sufficient to alleviate this.  However, assuming that the frequency 2nd 
derivative 
persists at -0.95$\times$10$^{-21}$ Hz s$^{-2}$, J0537 will spin at 31 Hz 
in $\sim$3,800 years, and may be a strong optical pulsar by that time.

As mentioned above, the single, narrow peak of the pulse profile
of J0537 is likely an indication that we observe its X-ray
pulsations at least partially (the pulse width is $\sim$10\% of
the period) as a result of the distance$^{-1}$ law, and its 2-10 
keV pulsed X-ray luminosity could be lower than
2$\times$10$^{35}$ erg s$^{-1}$ as estimated by 
\citep{Ma98}.

\subsection{The positron excess, WMAP `Haze', and Fermi `Bubbles'}
\label{Positrons}

A beam of protons, with kinetic energies of up to 2.2 GeV or 
greater, will eventually produce electrons with similar energies.
Electrons/positrons with even higher energies may result from photon 
scattering because the cross section varies inversely as the square 
of the particle mass,
and/or the leptons may, in turn,
be further accelerated by magnetic reconnection, in wound-up 
magnetic fields near the Galactic center, or other mechanisms
\citep{Sch09}, possibly to TeV
energies, to produce the WMAP `Haze'/FERMI `Bubbles' observed in that 
direction \citep{Fink04,Su10}.  
In addition, the SNR loss/relativistic
injection of $^{56}$Ni positrons into the pulsar jet,
which occurs because of the bipolarity of SNe, and which also makes
them unfit for easy cosmological interpretation, may show up as
an excess in cosmic ray data [153-155].  This may offer
a satisfying resolution for the apparent anomalous dimming of
distant SNe explained in terms of local cosmic ray abundances,
without invoking the creation of electron-positron pairs near
pulsars and their acceleration along `open' field lines which 
do not exist if one believes in the validity of the Sommerfeld 
equations \citep{KPR74,BR01}.  TeV fermions generated by SNe and/or
pulsars would alleviate the short lifetime problem ($\sim$10$^6$ yr)
associated with the Galactic center as a potential direct source.
SLIP does not require pulsars to generate positrons in order
for them to emit radiation, so the 10-100 GeV positron flux 
may not be dominated by contributions from local pulsars,
but may be provided instead by the supernovae that gave birth to 
them recently.  The main difficulties lie in accounting for how
and why the excess rises at energies above 10 GeV and below 100 GeV.

Protons at multiple GeV energies will destroy the (relatively
stationary) heavy nuclei they collide with, resulting in the 
release of free neutrons which can drive the r-process in supernovae.
A density of 10$^{24}$ cm$^{-3}$ in free neutrons has been invoked to 
drive this process, but SLIP can provide this near the center of 
the newly-formed remnant.  SLIP clearly also drives the r-process,
visible in early Ic spectra, at the poles of SNe Ia/c, as described 
in the Introduction, where the influence of annuli well outside of the 
stellar core bypasses that same core to produce focused beams at the 
two poles, where C and O are both available.
Detailed calculations are deferred
for future efforts (see, e.g., \citep{Mu11}).  

\subsection{The SS 433 jets}
\label{SS433}

The extreme collimation of the initial beam and jet in SN 1987A, 
and even the strong H${\alpha}$ seen in its spectrum near day 
19.6 are suggestive of a similar situation in SS 433
(e.g., \citep{Ma79}),
and a rapidly spinning pulsar has been proposed for
the underlying central engine (e.g., \citep{Ku99}), although
mass estimates for the compact object vary from 0.8 to
16 M$_{\bigodot}$ 
[159-162].  The big problem with
SS 433 is accounting for the mechanical power
necessary to produce $10^{34}$ erg s$^{-1}$ of
H${\alpha}$ from recombination in two jets with velocities
of 0.26 c, which could amount to 10$^{42}$ erg s$^{-1}$
if each atom has only one recombination.
Because of the few minute variability
of the H${\alpha}$ \citep{Gr81,Ma84} with no variability
of its velocity $>$1\%, the standard method of its solution 
is to invoke a clumpiness of 10$^5$ in the jets titled 
`bullets,' (e.g., \citep{Ch03}), so that each 
atom recombines many times ($\sim$10$^{2.5}$).  

Although \citep{Br91} constrained the parameter space for 
the bullets, assuming the H${\alpha}$ production mechanism
was an interaction of the jets with a B-star wind, the 
possibility of many excitations per atom, due to beamed 
radiation, was an admitted exception to the requirement
for bullets. 
Of course this is exactly what SLIP does, in the process
easily producing the kinematic and collimation properties 
of the SS 433 jets from a ms pulsar buried in hydrogen-dominated 
accretia, where L${\rm \infty}$ at the source is redshifted to 
L${\alpha}$ in the mature jets (this is exact for $\beta$=0.28,
but the actual $\beta$ is lower [0.26], consistent with the 
finite bandwidth necessary for the jet acceleration mechanism to 
work), the usual SN boost mechanism being unavailable because of 
the lack of high Z accretia.  

If SLIP turns out to be 
a necessary mechanism to produce the jets and their 
H${\alpha}$ flux, then the rarity of SS 433 may be less 
due to the wind from a Population I B star, than
to its association with a weakly magnetized ms pulsar,
which clearly did not originate from a massive, solitary
star similar to what we think describes the primary.
The best chance to detect a
pulsar in SS 433 may be during a low state, should that
ever happen, there having been no sign of any pulsar 
(e.g., \citep{Be79}) in the high state it has 
occupied since the discovery of its bizarre behavior in 
1978.  
A 2 ms pulsar could power the jets for 10$^4$ 
years if each atom produces $\sim$20 H${\alpha}$ photons, 
not even accounting for any spinup due to accretion.
It is more likely, however, that the spin period of SS 433
is larger because a 2 ms period is associated with the
evisceration of SN progenitors by SLIP, and,
in the presence of plasma at many $R_{\rm LC}$, as happens
in both SS 433 and SNe, will have a very large spindown, 
whereas the actual spin of SS 433 will settle to a rate 
where spinup from accretion is balanced by spindown due
SLIP.  This rate will likely be a bit higher than 
what is typical for the luminous LMXBs, perhaps closer to a 
period of 5 ms, because of the lack of heavy elements loading 
SLIP (see $\S$\ref{Recycling}).


\subsection{Star Formation Avalanche in the Early Universe}
\label{earlyU}

Computer simulations of clustering in the Early Universe fail 
without the addition of some extra agent to increase the rate 
at which clusters form (see, e.g., \citep{B98}).  The problem 
is that the agents which can be easily inserted into the codes, 
such as dark matter or dark energy, almost always involve 
controversial physics which may not be valid.  Aside from such 
agents, the only known mechanism which is frequent and violent 
enough to possibly play the same role are the pulsar-driven 
jets resulting from the supernovae of the first stars.

Although high velocity material will suppress star formation locally
(M.-M. Mac Low, 2011, private communication), even the early 
Universe (z$\sim$7) is vast, and this same material will initiate 
star formation [169-172] at some distance from the pulsar in the two 
initially polar jets.  Without spreading, the Mystery Spot column, 
${\sigma}_{\rm MS}$,
of 0.5 gm cm$^{-2}$ will penetrate even the early Universe at z=7, with 4\% 
of the present day critical density in baryonic mass.  
Assuming, for the sake of simplicity, a linear (conical) spreading 
scale of $\ell_0$ = 3 $\times$10$^{16}$ cm in an interstellar medium of 
density of ${{\rho}_{\rm B}}_{\rm 7}$=1.9$\times$10$^{-28}$ g
cm$^{-3}$ (4\% of critical density at z$\sim{\rm{500}^{1/3}}-1 \sim$7),
and solving for the pathlength, $\ell$, whereby a column of mass
equal to the spread-diluted Mystery Spot column has been swept up 
in its last 10\%, we get: 

\begin{equation}
\ell \sim 200~{\rm pc}~({ {\ell}_0 \over {3\times10^{16}} })^{2/3}
({{\sigma}_{\rm MS} \over 0.5})^{1/3} ({{{{\rho}_{\rm B}}_{\rm 7}} \over {1.9 \times 10^{-28}}} )^{-1/3}
\end{equation}

The collimation of the Mystery Spot is not exactly known, but it can't 
be as extreme as the 10$^{4\rightarrow5}$ of the initial beam/jet.  
However, it could easily amount to a few square light days at a distance 
of 10 light days.  In that case, the plume would be about a few light years 
across at 100 pc.  Along the way, shearing vortices may initiate star 
formation, with higher velocities at the larger distances from the endpoint.  

The actual situation may be much more complicated, as the decline 
of the jet collimation with time will impact the previously formed vortices.  
The stars formed by this process  
may also initiate another round of star formation as they move through 
the primordial ISM, and so on, with each cluster retaining a tiny, though
important fraction of the initial jet velocity.  Infant clusters in close 
proximity
with nearly matched velocities may merge due to their sustained mutual
gravitational attraction.  In this way, the SLIP jets may initiate a star 
formation avalanche, possibly leading to mini-cluster and/or stream formation 
in a few million years, with no requirement for dark matter.  The recent
discovery by GALEX \citep{OD11,Clavin} of
velocities, between close pairs of distant galaxies, which do not reflect
their mutual gravitational attraction, may be the result of the separation 
velocities of pulsar-driven jets in their formation of star clusters.  Again, 
detailed calculations are deferred for future efforts.

\subsection{Pulsar `recycling'}
\label{Recycling}

Pulsars with rapid spins and weak magnetic fields are commonly
referred to as having been `recycled,' wherein these attributes
are a result of a (usually) long dead pulsar undergoing an 
extensive interval of accretion from a binary companion which
it acquired or already had, that as a result has increased its 
spin rate from the few second period typical of a 
very old strongly magnetized pulsar, to the ms period range,
and as a side effect either of the accretion, or slow decay of the 
effective magnetic dipole by some other mechanism because of  
advanced age, or both, reduced its original several Teragauss magnetic 
field by 2 to 4 orders of magnitude.  

Here, however, we have seen evidence that the spindown due to 
SLIP can greatly exceed any spinup due to accretion.
Thus, the only way to tell if a pulsar has been recycled is by a mass
substantially in excess of 1.3 M$_{\bigodot}$ (as is the case for 
J1903+0327),\footnote{Given that the best estimate for the mass of
Her X-1 is 1.30 M$_{\odot}$ \citep{MN76,M83}, and the likelihood that it 
has been accreting 10$^{-9}$ M$_{\odot}$ from HZ Her for several hundred million 
years (with gaps -- \citep{JFL73}), the best estimate for the standard, initial strongly
magnetized neutron star mass is likely closer to 1.25 M$_{\odot}$, which 
is supported by the masses of the strongly magnetized pulsars, J0737-3039B at 
1.2489(7) \citep{K06}, and J1906+0746 at 1.248(18) M$_{\odot}$ 
\citep{Kasian,LSM06}, respectively (see also \citep{KKT10} and references 
therein).  This could mean that the standard Chandrasekhar mass is closer to 
1.35 M$_{\odot}$, unless it is actually higher for {\it weakly} magnetized
progenitors, or that 0.05 M$_{\odot}$ is typically excluded from the
core-collapse for strongly magnetized neutron stars.} 
but even then, there's no guarantee that the pulsar has 
suffered any net spinup, the result depending on accretion rate
and history, spin rate, and magnetic field strength and configuration.  

\begin{figure}
   \centering
   \includegraphics{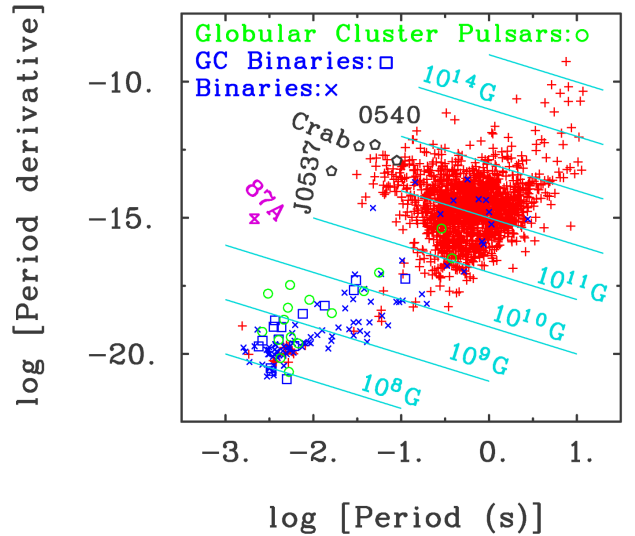}
   \caption{
   A plot of 1723 pulsars from the ATNF Pulsar Database in 
   the log period-period derivative plane.  Non-binary,
   non-globular cluster pulsars are marked by `+' signs,
   non-globular binary pulsars by $\times$'s, non-binary
   globular cluster pulsars by circles, globular binary
   pulsars by squares, four strongly magnetized, young 
   pulsars by pentagons, and the 2.14 ms signal from 
   SN 1987A by an hourglass.}
              \label{ppdot}%
    \end{figure}

To gain further understanding of this process, 1723 pulsars are located on 
the log period-period derivative plane in Figure \ref{ppdot}.  On this plot 
the connection between the `ordinary' pulsars  in the upper right to the 
`recycled' pulsars in the lower left is a two-lane, one-way highway.  
The top lane (higher period derivative) contains all six of the globular
cluster pulsars, and nine Population I pulsars, for the period range from 
10 to 100 ms, while the bottom lane contains {\it just} Population I 
pulsars for that range.  Metalicity and gravitational contamination are
unlikely contributors to this pattern.  The lanes have numerical slopes 
of about 1.5, meaning if pulsars migrate to the right and slightly up 
in the lanes, the period derivatives gain 3 parts for every 2 parts gain 
in period.  By contrast the same ratio is 3 to 1 for the young (4,000-year 
old), 16.1 ms PSR J0537-6910 (see $\S$\ref{0537}), twice the slope of the 
two lanes (its effective magnetic dipole increases by 0.02\% per year), so 
perhaps the pulsars in these lanes actually do migrate right and up along 
their respective, sloped paths.  

Unfortunately, none of the six globular cluster 
pulsars in the upper lane within that period range have measured second 
frequency derivatives (which should be negative for this migration but 
which will also suffer extremely from gravitational contamination), so 
we can't test the hypothesis that these pulsars are migrating on the 
slightly upward path.  Of the 32 non-globular pulsars in the lanes within 
this period range, 2 have negative and 2 have positive frequency second 
derivatives, so half of these four pulsars are not migrating within the 
lanes.  With a period derivative of 1.88e-20, J1903+0327 lies on the low 
boundary of the short period end of the upper lane.  The offset between 
the two lanes corresponds to nearly two orders of magnitude in period 
derivative, but only one order of magnitude in field strength.

Still, this offset between the two lanes is likely to be
due to different accretion histories, provided accretion
does in fact reduce pulsar effective dipoles, as postulated 
by \citep{CR93} and \citep{Rud95}.  The simplest 
interpretation is that pulsars in the upper lane, containing 
about ten of eleven globular cluster pulsars with periods $>$8 ms, 
have never suffered accretion and were born fast and weakly
magnetized.  

It is even difficult to argue in favor of a strongly
magnetized origin for those pulsars in the lower lane, as
it would seem strange that this population starts only from
those `ordinary' pulsars with the smallest period derivatives, and
neither from the much larger population of `dead' pulsars with even
lower period derivatives.\footnote{This assumes that the accretion cycle 
from these to the lower left is always hidden, i.e., the accretion is 
never interrupted until the pulsars arrive near the pileup near 5 ms, 
nor the `normal' population with higher period derivatives.  We reject 
the hypothesis that this does happen during the accreting phase but a 
large fraction somehow fall neatly into this well focused lane once 
their accretion ceases.}
It may be more likely that this (lower) lane is from faster pulsars 
which accreted from binary companions to become LMXB's that have 
ceased to accrete, and are left there.  
Similarly, the upper lane may populate the main group just above its
lowest point.  




So on average these pulsars slowed during their phase of high 
accretion, which is in line with the modest QPO frequencies seen in 
many LMXB's,\footnote{As well as the 104 Hz optical signal observed 
from the LMXB 4U1728-24.} 
and the high frequencies seen from most of the accreting millisecond 
X-ray pulsars, which have very low accretion rates.



As the accretion rate diminishes
when the binary companion begins to withdraw from its Roche limit,
for whatever reason, as long as this is a gradual process, the pulsar will 
have an era of spin-up, possibly resulting in the pileup at 3-5 ms periods.  
Here however, pulsars are found with period derivatives still lower than 
those of the lower path, thus requiring a population of neutron stars 
which were born only weakly magnetized, and likely spinning in the 
millisecond range.


Thus these two discrete lanes connecting the ``normal'' pulsar
group in the upper right to the ms pulsars in the lower left 
require a source of pulsars whose spin rates are faster than 200 Hz.  
Again, the most likely source to populate these paths is the source 
of born-fast 2 ms pulsars from the DD process.  Since there are 
orders of magnitude more neutron stars generated by the DD mechanism 
than by Fe catastrophe, there should be a much larger reservoir of 
`dead' weakly magnetized pulsars than their `dead' strongly magnetized 
cousins.  A study of the high space 
velocities of solitary neutron stars as compared to the masses and low 
eccentricities of double neutron star binaries has also reached the 
conclusion of the necessity of two mechanisms to produce neutron stars 
\citep{vdH10}, the DD process being the second (though overwhelmingly
predominant) mechanism.

\section{Conclusion}
\label{conc}

We have derived a self-consistent solution for the onset
($\sim$11 $\ell$t-d), and depth ($\sim$13 $\ell$t-d),
of the polar ejecta of the progenitor of SN 1987A,
the energetics of its beam/enhanced UV flash ($>$10$^{40}$ erg
s$^{-1}$), the kinetics of its jet ($\beta \sim 0.95$ for 
$\sim$10$^{-6}$ M$_{\bigodot}$), and angle from the line of sight 
to the Earth ($\sim$75$^{\circ}$).  There is plenty of
evidence for the absence of any strongly magnetized pulsar
within SN 1987A, and such a pulsar may not have the rotational
energy to account for the kinetics derived anyway.
Thus, we have argued, through the paradigm of a model of pulsar
emission (SLIP) from polarization currents updated supraluminally
beyond the pulsar light cylinder \citep{Ar98},
that SN 1987A, its beam/jet, `Mystery Spot,' and possible
2.14 ms pulsar remnant, are intimately related to SS 433, Sco X-1 
and as many as 99\% of LMXBs, GRBs, ms pulsars, and other SNe, 
including all Type Ia SNe.  The SLIP model explains, in a 
natural way (Eq.~1), the changes over time observed in the 
collimation of the SN 1987A beam and jet.

The time lags, energetics, and collimation of $\ell$GRBs are
consistent with those of 1987A's initial beam, jet, and the
inferred polar ejecta.  When 
the bipolarity of SN 1987A is interpreted through this model, 
its pulsar clearly had ablated part of the $\sim$10 M$_{\bigodot}$
of ejecta, eviscerating the remnant by blowing
matter out of its poles at speeds up to 0.95 c or greater,
and enforcing a toroidal geometry on the remaining
equatorial ejecta.  Since there is no reason to suggest that
this is not universally applicable to all SNe, this geometry
has grave implications for the use of Type Ia SNe as standard
candles in cosmology (see,.e.g., \citep{WT05}, and 
$\S$\ref{Ia/c}).

Lines observed from Type Ic SNe provide proof that SLIP beams
drive the r-process at the rotational poles, in or near the
progenitor photosphere ($\S\S$\ref{intro}, \ref{Ia/c},
\& \ref{Positrons}).  No other model of SNe can explain
why such lines are only observed in SNe Ic (which are 
actually SNe Ia when obeserved away from the poles, and
also when the r-process lines can not be observed), the
reason being that C and O are only available at the surface 
of SNe Ia/c.
 
Because even weakly magnetized neutron stars buried within plasma 
can spin down at a rate which exceeds their ability to spin up from 
accretion by orders of magnitude, pulsars interacting with the rest
of the remnant progenitor (if this rest exists), clearly can not
be ignored -- they are the driving force behind 
the disruption of their progenitor stars in supernova explosions, in 
spite of the difficulty of detecting them within the remnant, now 
explained by the associated beaming.  This has observable consequences:
the start of the associated UV flash will show up within minutes, or even 
seconds, of the core-collapse and neutrino flux, and the timetable of many 
SN 1987A events needs to be revised accordingly.  In addition, the
existence of a robust mechanism for disrupting progenitor stars means
that the stellar mass at which core-collapse continues on to create
a black hole is much larger than previously thought -- perhaps closer
to 75 M$_{\odot}$ or even higher.

There appears to be no need to invent exotica
to explain GRBs -- the SLIP model provides the young pulsar
(or even a near-Chandrasekhar-mass white dwarf) with an initially 
tightly collimated beam as the central engine, and makes
the very specific and testable prediction that GRB afterglows are,
in fact, pulsars, usually spinning at 500 Hz in the proper frame.  

Beaming has always been a wild card for pulsars, and for the
SLIP model, the beaming is effectively infinite -- part of the 
radiation decays only as the first power of the distance, 
and the two cones of this beam (which subtend a solid angle of
measure 0) will change in polar angle as plasma
conditions change outside of the pulsar light cylinder.  Beams
and jets which are less than maximally collimated (those which 
originated from annuli of less than maximal radii) may not have 
enough concentrated energy to break through extensive surrounding
material, and rather help to lift it away from the neutron star.
Rotating objects for which pulsations are not detected may simply 
be beamed in other directions.  Thus the disappearance, or 
persistent absence, of pulsations from rotating compact objects, 
such as those which might lie within the Cygnus Loop ($\S$\ref{Plasma}), 
SN 1987A ($\S$\ref{early}), SS 433 ($\S$\ref{SS433}), and Sco X-1 
($\S$\ref{SCOX1}) and other LMXBs ($\S$\ref{LMXBs}), does not 
automatically imply that there never was a pulsar, and/or that 
the compact object is a black hole, the energy requirement for a pulsar 
being negligible because of the distance$^{-1}$ law.  Finally, it may 
be that all transient events observed in the very distant Universe, 
including the outbursts 
of QSOs, are seen because they have a similar inverted distance law, 
which is a result of a supraluminal excitation of one sort or another.

Such restrictive 
beaming ($\theta_{\rm V} \ll 1$ and plasma at many $R_{\rm LC}$) 
explains, in a natural way, the non-detections of pulse frequencies 
from the luminous LMXB's, and less restrictive beaming ($\theta_{\rm V} 
\cong 1$ and plasma at few $R_{\rm LC}$) the intermittent detections 
of oscillations from the less luminous LMXBs -- the accreting 
millisecond X-ray pulsars.  The neutron star components of the LMXB's
will settle to a spin rate where spinup from accretion, which has an
absolute limit, balances spindown
from SLIP losses, which increases with mass transfer rate
almost without limit.  Thus the neutron star spin rate in LMXBs will
vary inversely with luminosity (see $\S\S$\ref{LMXBs} \&
\ref{Recycling}).  

As is the case for many other things, the Devil in this Universe may
lie in the details.  Thus while the most extensive simulations of the
early Universe still require dark matter to form galaxies (see. e.g.,
\citep{He05} and \citep{UW04}), the jets 
driven by the pulsars resulting from the SNe of the first stars may
provide the perturbations necessary to initiate the next generation
of star formation in such a way as to include much of this generation
in rapidly growing star clusters and/or streams (see $\S$\ref{earlyU}).
Thus, although it may 
take longer to form the first stars without dark matter, the formation 
of clusters via pulsar-driven jets may make up this time.  The peculiar 
motions observed from distant galaxy pairs \citep{OD11,Clavin}
which would otherwise be expected to show motions 
reflecting their mutual gravitational attraction, 
may be due to the pulsar jet-star-formation process.

In addition, because of the large systematic effects involved in 
measuring Type Ia (and all other) SN luminosities due to the 
unsuitability of their geometries 
(see $\S$\ref{Ia/c}, \citep{Be05}, and \citep{Kr05}), 
we simply can not tell if they
suffer any anomalous dimming with cosmological distances, without
further understanding of the SN process {\it and the ability to
apply this understanding to carefully select the distant sample}.
This was the only really direct evidence that the expansion 
of the Universe is accelerating, and now the scientific
community must again question the existence of dark energy.  Without 
it, in addition to the possible lack of need for dark matter in the 
formation of star clusters, there may be no longer any numerical 
coincidence to support the role of dark matter in Concordance Cosmology.  
Recent observations of solitary ring galaxies have 
also cast significant doubt on the existence of dark matter 
\citep{NP07,MNP09}.

Thus, aside from bizarre entities such as quantum entanglement 
currents, there is not much evidence for any dark `stuff' present in 
the recent Universe (z$\le$1) having any observable effect, and this 
has been due in significant part to a new understanding of the 
role of pulsar-driven jets.
Still, the evidence is strong that $\Omega_{\rm total} = 1$ at the era
of recombination ($z \sim 1000$ -- e.g., \citep{Sp03}).  
However, in addition to kinematic redshift, gravitational redshift
becomes important when looking back that far, and
our knowledge of gravity is incomplete at best.  Perhaps the best
assessment of this apparent factor of $\sim$20-25 difference in 
$\Omega_{\rm total}$ between the era of recombination and the
more recent Universe is that it remains a deeper mystery.

The path to the intimate and productive study of the SN and GRB processes 
is now clear: with high time resolution observations of GRB afterglows, 
through the selection effect of the inverted distance law operating across 
billions of light years, we let the Universe reveal the behavior of the 
pulsars responsible.  It might appear that a Universe without the 
need for dark matter or energy, collapsars, hypernovae, pair instability 
SNe, super-Chandrasekhar mass white dwarfs, frequent collisions 
of massive stars, and neutron star-neutron star mergers which 
make sGRBs, is much less `exotic' than previously thought. 
However pulsars, i.e., clocks and minutes-old neutron stars to boot, 
which can be seen across almost all of this same Universe 
(which is never in equilibrium), including 
even core-collapse remnants of the first stars, and which 
may yield practically instant redshifts just from their pulse
frequencies alone, may well suffice in explaining all of the 
issues which gave rise to the previously mentioned entities, 
are more in line with Occam's Razor, and are also, of themselves, 
extremely worthy of study.

\vskip 0.15in
I would like to thank Dr.~John Singleton for supporting this
work through the Los Alamos National Laboratory LDRD grant 20080085DR, 
``Construction and Use of Superluminal Emission Technology Demonstrators 
with Applications in Radar, Astrophysics and Secure Communications,'' and
grant 201100320ER, ``Novel Broadband TeraHertz Sources for Remote 
Sensing, Security and Spectroscopic Applications,'' as well as Drs.~Joe Fasel,
Todd Graves, Bill Junor, and Andrea Schmidt.  I am also grateful to Larry 
Earley of Los Alamos, Jim Johnson and Carl Pennypacker of Lawrence Berkeley
Laboratory, and John Saarloos for loans of equipment for use at Lick 
Observatory, as well as Keith Baker, Kostas Chloros, Elinor Gates, Bryant Grigsby, 
Rem Stone, and the rest of the entire staff of of Lick Observatory for their assistance,  
and M.~T. ``Brook'' Sanford of Los Alamos for programming assistance.  I would also
like to thank Joel Roop and Andy Toth of Keithley Instruments for the loan of
an engineering unit of the KUSB 3116 for testing at home while far away from
Lick Observatory.  This work was performed under the auspices of the Department of 
Energy.

\clearpage

\section{Appendix: A short SLIP primer}
\label{Appendix}

Models of pulsar emission usually involve the motion of charged particles
between the neutron star and the light cylinder.  The Outer Gap model 
\citep{CHR86} produces the emission by particle acceleration in the gap just 
outside of the last closed magnetic field line within the light cylinder and 
the null surface (${\bf \Omega \cdot  {\rm \bf B}} = 0$).  The Polar Cap models
involve pair production, electron acceleration, and production of secondaries
close to the magnetic poles of the neutron star \citep{P11}.  Other models
produce emission via inverse Compton upscattering of lower energy photons
(see, e.g., \citep{PD11}, and references therein).
In the SLIP model radiation from polarization currents induced
beyond the light cylinder initially focuses on the light cylinder and
drives a particle wind along with a beam.  SLIP is indifferent to material
within the light cylinder, and for isolated pulsars relies on material
from the ISM concentrated by the gravitational field of the neutron star
(see, e.g., \citep{Ar09}).

Electromagnetic radiation can be generated by a time-dependent polarization,
${\rm {\bf P}(t)}$, just as it can from currents of free charges, ${\rm {\bf j}}$,
displacement, ${\rm {\bf D}}$, and electric field, ${\rm {\bf E}}$ \citep{JDJ}:
\begin{equation}
\nabla\times{\rm {\bf H}}={{4\pi} \over c}{\rm {\bf j}}+
{1 \over c}{{\partial {\rm {\bf D}}} \over {\partial t}} = 
{{4\pi} \over c} ( {\rm {\bf j}} + { {\partial {{\rm{\bf P}} } \over {\partial t} }})+
{1 \over c}{{\partial {\rm {\bf E}}} \over {\partial t}},
\end{equation}
where ${\rm {\bf H}}$ is the magnetic field and $c$ is the speed
of light in vacuo.  Although the motion of the charged particles in the
polarization currents is, in general, slow, there is no limit to the 
speed at which they can be modified, similar to a wave produced by
fans in a stadium at a sports event.  In particular, the action of a
rotating dipole, such as a neutron star with a non-aligned magnetic
field, can change polarization currents much faster than the speed of
light at many light cylinder radii, $R_{\rm LC}$.

As the (mostly dipole) magnetic field of the neutron star rotates,
much of the polarization current field beyond $R_{\rm LC}$ will be
instantaneously parallel because the magnetic field in opposite
directions with respect to the axis of rotation is also opposite.
Thus the polarization currents can be thought of as a single
vector function of radius, which is rotating rigidly with with
the pulsar rotation frequency, $\omega$.


   \begin{figure}
   \centering
   \includegraphics{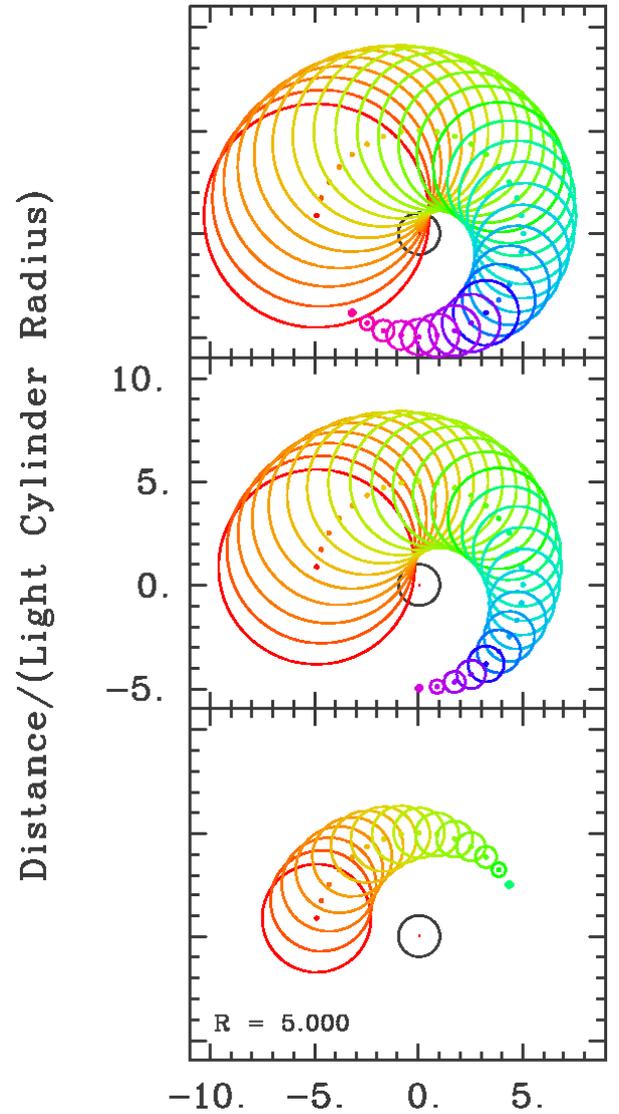}
   \caption{
   (Bottom to top) A progression of Huygens' wavelets for a 
   supraluminal source moving at 5 times the speed of light,
   with 36 wavelets in a full circle.  The light cylinder
   footprint is marked by the small, black circle.}
              \label{super5.00}%
    \end{figure}

The influence of currents from an annulus of a given radius, on a 
point on the light cylinder and in the same plane perpendicular to the 
axis of rotation, will then consist of the magnetic field produced by 
such currents, and the influence of the cyclotron radiation and strong 
plasma turbulence induced by the currents.  However, the influence of the
currents is affected, in a non-trivial way, by their centripetal 
acceleration in the supraluminally rotating frame of reference
(see, e.g., \citep{Ar03}), a complication which we will
ignore in this discussion.
The net influence on the point can at least be understood by the 
influence of a progression of a series of Huygens' wavelets, as shown in 
Fig.~\ref{super5.00}, which shows many wavelets intersecting a point
tangent to the light cylinder.  Ultimately the wavelets from nearly a 
third of a whole revolution will intersect at such a point still further
clockwise around the light cylinder.  
More than this would be 
unproductive, both because directions of further contributions to 
${\rm {\bf H}}$ at the point would be more than 60$^{\circ}$ away from 
the mean phase, and, of course, the induced time-varying cyclotron radiation
or strong plasma turbulence would also suffer out-of-phase contributions.

   \begin{figure}
   \centering
   \includegraphics{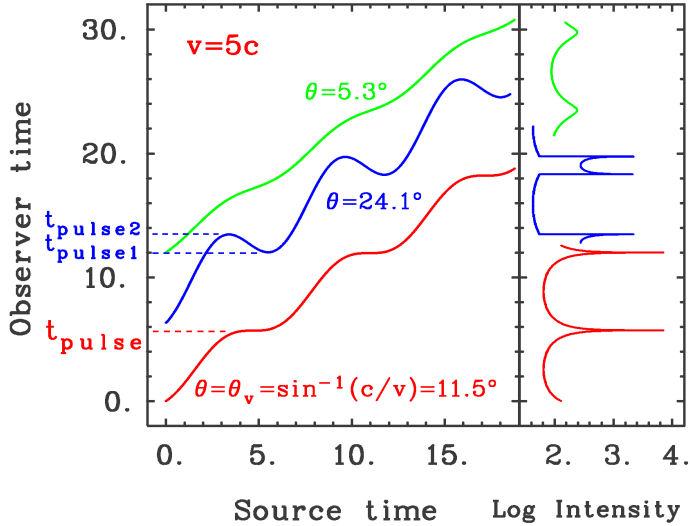}
   \caption{
   (Left) Observer time as a function of source time for a supraluminally
   induced circular excitation at 5 times light speed with a period of
   2$\pi$ for three polar 
   directions.  Delays in the observer time of order millions have been 
   suppressed, and the three curves have been arbitrarily offset by a few
   units for clarity. (Right) The integral, into 512 discrete bins per 
   cycle of observer time phase, of source time dwell versus observer time
   for the three curves (i.e., pulse profiles).  }
              \label{round}%
    \end{figure}

As time progresses further, this focus point, or `cusp', will split as it
moves out of the plane, spiralling out on two cones of polar half angle given 
by Eq.~(1) (see, e.g., $\theta_{\rm P}$ of 
Eq.~(13) of \citep{Ar07}).  The dependence of observer time on source 
time for three different $\theta_{\rm P}$, and the three different resulting
pulse profiles, are shown in Fig.~\ref{round}.  The 
singly-peaked pulse profile (repeated once) for the direction corresponding to 
the `cusp,' for this $v=5c$ case (polar angle of $\sim$11.5$^{\circ}$) is shown
in the lower part of the right frame.  Almost exactly half of the $\sim$1,000
pulsars in the Parkes Multibeam Survey have a single pulse peak with a full width 
at half maximum less than 3\% of the pulse period (see e.g., \citep{Man05},
and $\S$\ref{link}), and SLIP predicts that all of these will be very closely
spaced doubles (see $\S$\ref{HistR}).

For larger polar angles
the pulse profiles are obviously double, and less sharp (right frame, middle), 
and for smaller polar angles (right frame, top), the pulse profiles are again 
single, but still smaller, and eventually not sharp at all.  The advantage in 
intensity of one pulse profile over another will depend on how many points are 
used in plotting the curves -- the more points, the greater the advantage, 
until the limiting case of the distance$^{-1}$ law is reached.
The point of inflection in the lowest curve of Fig.~\ref{round}
is the circular analog of a shockwave, and it is this that results in the
distance$^{-1}$ law.  

If the rotation-powered pulsars which radiate from just beyond the light
cylinder (see $\S$\ref{0537}), produce the characteristic singly-peaked 
pulse profile, and a beam which, according to Eq.~(1), is very nearly 
equatorial, how is it that pulsars with doubly-peaked profiles find the 
phase space in greater polar angle to generate those profiles?  

One possible production mechanism for complex pulse profiles is a 
multipole component of the neutron star magnetic field.  Barring that, 
the Crab pulsar may offer a clue to this conundrum.  Its inter-pulse, which
alone is similar to the small polar angle top pulse profile of Fig.~\ref{round},
shows the emission bands in the 2-10 GHz range predicted by SLIP, while its
main pulse does not \citep{HE07}.  
The possibility that the main pulse is a result of a less homogeneous 
environment and the bands are smeared out to uniformity seems to be excluded 
by these observations.  Thus the Crab's main pulse may arise
from a different location, i.e., a different radius, which 
would also imply emission from a radius greater than just outside the light 
cylinder, which in turn could mean plasma sheets generated closer to the pulsar
travelling outward through $R_{\rm LC}$.
Still, the equal 180$^{\circ}$ swing in polarization across both the main and
interpulse supports a SLIP process for both.

The movie of the Crab \citep{He02} in Chandra X-rays appears to show at least  
one nearly equatorial fan of pulsations and 0.5 c wind, 
as predicted by SLIP through Eq.~(1) for material just outside the light 
cylinder, with our line of sight between the lower rotation axis and the lower 
near-equatorial wind component.  Weisskopf et al.~\citep{We12} interpret this 
as a single
circular feature projected onto the plane of the sky, and offset from the
plane of the sky containing the pulsar as a fan beam of some 4.5 degrees off
the equatorial plane of the pulsar, which would indicate that
the location of the SLIP emission occurs at 1.003 times the light cylinder
radius.  It is not clear if there is another fan, as predicted by SLIP, which 
does not appear because its aspect produces less back-scattering.
The width of the bottom lip of the `bell' supports a two-fan interpretation 
of some sort, and may imply that plasma around the Crab pulsar does not extend 
a large fraction beyond $R_{\rm LC}$ ($\sim$1,600 km), although it has to 
exist more densely at great distance to reveal the very distant dominant ring.

Because our line of sight to the pulsar is much closer to the pole than the
fan in the opposite hemisphere from that which produces the circular arc, 
the Crab's interpulse may be a small polar angle feature similar to the pulse 
profile in the upper right of Fig.~\ref{round}.  There is the possibility that 
plasma is trapped at a few $R_{\rm LC}$ and leads to an emission fan that {\it is} 
more nearly polar, but just not easily visible in the movie, and could be associated 
with a second arc visible to the right (west) of the innermost arc.  The fan in the 
opposite hemisphere to the one that produces this second arc may actually lie at, or 
closer, to the pole than our line of sight, and this would ease the angular requirement 
for producing sharp and double pulses, and may, in some way, reveal the production 
mechanism of the Crab's main pulse.

The movie also shows the Crab has polar jets, and there are {\it two} possible sources 
of supraluminal excitations of the same concentric feature which reveals the
fan of wind and pulsations.
The first supraluminal excitation may be due to the discrete waves of particles, 
made famous by the movie, impacting the concentric features.  Although the timing of 
the particle wave arrivals at all azimuths will not be exact, it may be close enough 
to drive substantial polar jets.  The second supraluminal excitation is due to the 
actual $\sim$30 Hz pulsations themselves, and for these the timing will be exact
because there will frequently be enough material in the features to trace the path 
of each individual pulse over a wide range of azimuth, in spite of the fact that they 
are of order 100s of millions of light-pulse periods away (a few light-months).  

If pulsars on the `cusp' have to be {\it very} close to satisfying
Eq.~(1), it may not take a lot of range in polar angle closer to 90$^{\circ}$
to populate the doubly- or multiply-peaked sample.  Those pulsars with 
smaller polar angle may be relatively suppressed due to the poor detectability
of their bland, singly peaked pulse profiles (the top right curve in 
Fig.~\ref{round}).  Further studies of artificial pulsar populations may 
yield a definitive answer to this question about the distribution of types of
pulse profile.

\clearpage
\addtocounter{section}{+1}
\section*{}


\end{document}